\documentclass[%
reprint,
amsmath,amssymb,
aps,
prd,
]{revtex4-2}

\usepackage{graphicx}
\usepackage{dcolumn}
\usepackage{bm}
\usepackage{hyperref}

\usepackage{xcolor} 

\definecolor{darkblue}{rgb}{0.0, 0.0, 0.55} 
\hypersetup{
	colorlinks=true,
	linkcolor=darkblue,  
	citecolor=darkblue,  
	urlcolor=darkblue,   
}

\renewcommand{\eqref}[1]{%
	\hyperref[#1]{(\ref{#1})}%
}

\begin{document}
	\preprint{APS/123-QED}
	
	\title{Cosmological Tensions in a Gauge-Invariant Modified Gravity}
	
	\author{Zahra Tabatabaei}
	\email{z\_tabatabaee@physics.sharif.edu} 
	
	\author{Sohrab Rahvar}
	
	\author{Shahin Rouhani}
	
	\affiliation{Department of Physics, Sharif University of Technology, P.O. Box 11365-9161, Tehran, Iran}
	
	\date{\today}
	
	\begin{abstract}
		In this work, we investigate a gauge-invariant formulation of modified gravity (GIMOG) wherein the gravitational interaction emerges dynamically from a scalar field following a first-order phase transition. This framework offers a unified cosmological history: it naturally generates a pre-inflationary phase, smoothly recovers the standard radiation and matter-dominated eras, and accounts for late-time cosmic acceleration without the need for a cosmological constant or dark energy. We evaluate the phenomenological viability of the model by confronting it with observational data across distinct cosmological epochs. At late times, the model is constrained using Pantheon+ Type Ia supernova data. In the early Universe, we impose bounds from Big Bang Nucleosynthesis (BBN), specifically utilizing the primordial $^{4}\mathrm{He}$ abundance. Our analysis reveals a distinct phenomenological tension: while late-time observations favor a stronger effective gravitational coupling, BBN constraints tightly restrict early-universe deviations from general relativity. We demonstrate that reconciling these constraints requires a smooth time variation of the effective gravitational constant, $G$, establishing a clear theoretical target for future precision cosmological tests.
	\end{abstract}
	
	\maketitle
	
	\section{Introduction}
	\label{introduction}
	The standard cosmological model, $\Lambda$CDM, successfully describes a wide range of astronomical observations, yet it relies on two enigmatic components: dark matter and dark energy. The evidence for dark matter is extensive, originating from the flat rotation curves of spiral galaxies \cite{rubin1970rotation, rubin2000one, sofue2001rotation}, the dynamics of galaxy clusters \cite{carlberg1996galaxy}, and gravitational lensing patterns \cite{bartelmann2001weak}. These phenomena consistently indicate a gravitational influence far exceeding that of visible baryonic matter, suggesting a universe dominated by a non-luminous substance. Despite decades of searching for particle candidates like WIMPs and axions \cite{arcadi2025waning, kuster2007axions}, no definitive detection has been made, leaving the nature of dark matter one of the most significant open questions in modern physics.
	
	An alternative paradigm suggests that the missing gravity problem is not due to undiscovered matter, but rather to a breakdown of General Relativity (GR) on large scales. This has motivated the development of modified gravity theories. One of the earliest and most influential is Modified Newtonian Dynamics (MOND), which alters gravitational laws at low accelerations to explain galaxy rotation curves without dark matter \cite{milgrom1988use}. While successful on galactic scales, MOND and its early relativistic extensions have struggled to simultaneously account for the dynamics of galaxy clusters and the precise measurements of the Cosmic Microwave Background (CMB).
	
	Among the more comprehensive relativistic alternatives is Modified Gravity (MOG), a scalar-vector-tensor theory that extends GR with a massive vector field and two scalar fields that dynamically alter the strength of gravity \cite{moffat2006scalar}. MOG has shown remarkable success in fitting flat galactic rotation curves and the mass profiles of galaxy clusters without invoking dark matter \cite{moffat2013mog, moffat2014mog}. However, it faces its own challenges, including discrepancies with gravitational lensing observations \cite{rahvar2019propagation, rahvar2022hamiltonian} and a significant theoretical inconsistency: the lack of gauge invariance in its vector field formulation.
	
	To address this fundamental issue, a gauge-invariant formulation of MOG (GIMOG) was recently developed, grounding the theory in a spontaneous symmetry-breaking mechanism \cite{rouhani2024mog}. This refined framework not only restores gauge symmetry but also yields novel cosmological implications, chief among them an effective gravitational constant, $G_{\text{eff}}$, that emerges from the vacuum expectation value of a scalar field. The development of a consistent, gauge-invariant theory opens a new window to perform rigorous cosmological tests.
	
	In this work, we confront the GIMOG framework with key cosmological observations from two distinct epochs: the late-time and early universe. We use the Pantheon+ dataset of Type Ia supernovae and cosmic chronometer data to constrain the model's parameters during the recent era of cosmic acceleration. We then derive independent constraints from the primordial abundance of helium-4, a sensitive probe of the expansion history during Big Bang Nucleosynthesis (BBN). Our central goal is to determine if a single set of GIMOG parameters can consistently describe the universe's evolution at both early and late times. Our analysis reveals a significant tension between the two datasets, suggesting that late-time observations favor a stronger gravitational interaction than is permitted by primordial nucleosynthesis.
	
	This paper is structured as follows. In Sec.~\ref{sec:theory}, we detail the theoretical framework of GIMOG and derive the modified cosmological field equations. Section~\ref{late universe} presents the analysis of late-universe data from supernovae and cosmic chronometers. In Sec.~\ref{sec:constrain}, we explore the early-universe constraints from BBN. In Sec.~\ref{sec:multi_epoch}, we perform a joint likelihood analysis to quantify the tension between the early- and late-time datasets. Finally, in Sec.~\ref{sec:conc}, we discuss the implications of our findings and conclude the paper.
	\section{Theoretical Foundations}
	\label{sec:theory}
	\subsection{Gauge-Invariant Field Equations and the FLRW Metric}
	Our cosmological framework is the spatially flat Friedmann-Lemaître-Robertson-Walker (FLRW) metric, with a metric signature of $(-, +, +, +)$. This choice is well-supported by cosmic microwave background (CMB) observations, which indicate a nearly flat geometry for the universe. The theoretical basis is the Gauge-Invariant Modified Gravity (GIMOG) theory, defined by the total action:
	\begin{equation} 
		S = S_g + S_v + S_s + S_M,
		\label{eq:s}
	\end{equation} 
	which incorporates tensor, vector, scalar, and matter field contributions. The gravitational action, $S_g$, is a modified form of the Einstein-Hilbert action:
	\begin{equation} 
		S_g = \frac{1}{16\pi} \int \frac{R}{G} \sqrt{-g} \, \mathrm{d}^4x,
		\label{eq:s_g}
	\end{equation} 
	where $R$ is the Ricci scalar. Here, $G$ is not a fundamental constant but a dynamical gravitational coupling determined by the vacuum expectation value of a complex scalar field $\phi$, such that $\langle |\phi|^2 \rangle = G$. The non-zero value of $G$ arises from a spontaneous symmetry-breaking process at low cosmic temperatures \cite{rouhani2024mog}.
	
	The vector and scalar components of the action are given by:
	\begin{equation} 
		S_v = \int \left( -\frac{1}{4}B^{\mu\nu} B_{\mu\nu} - \kappa A_\nu J^\mu \right) \sqrt{-g} \, \mathrm{d}^4x,
		\label{eq:s_v}
	\end{equation} 
	and
	\begin{equation} 
		S_s = \int - \gamma \left( \overline{D^\mu \phi} D_\mu \phi + V(|\phi|^2) \right) \sqrt{-g} \, \mathrm{d}^4x.
		\label{eq:s_s}
	\end{equation} 
	In these expressions, $B_{\mu\nu} = \partial_{\mu}A_{\nu}-\partial_{\nu}A_{\mu}$ is the vector field strength tensor, and $D_\mu \phi = \partial_{\mu} \phi + i w A_\mu \phi$ is the gauge-covariant derivative, with $w$ being the charge of the scalar field.
	
	The symmetry breaking is driven by a potential analogous to the Coleman-Weinberg mechanism, which allows for a first-order phase transition in the early universe:
	\begin{equation} 
		V(|\phi|^2) = m^2 |\phi|^2 + \theta |\phi|^4 + \frac{b}{3} |\phi|^6.
		\label{eq:v}
	\end{equation} 
	The parameters $m^2$ and $b$ are positive constants. The evolution is controlled by the reduced temperature $\theta = (T - T_c)/T_c$. In the low-temperature regime ($T \ll T_c$), the scalar field settles into a true vacuum, and the effective gravitational coupling takes the form \cite{rouhani2024mog}:
	\begin{equation} 
		G = \frac{-\theta + \sqrt{\theta^2 - m^2 b}}{b}.
		\label{eq:G_theta}
	\end{equation}
	
	\subsection{Field Equations and Cosmological Dynamics}
	Varying the total action \eqref{eq:s} with respect to the metric $g^{\mu\nu}$ yields the modified Einstein field equations:
	\begin{equation} 
		R_{\mu \nu} - \frac{1}{2} R g_{\mu \nu} = 8 \pi G \left( T_{\mu \nu}^{(v)} + T_{\mu \nu}^{(s)} + T_{\mu \nu}^{(M)} \right).
		\label{eq:eq_fi}
	\end{equation} 
	
	The energy-momentum tensor for a given component is defined as $T_{\mu \nu} = (-2/\sqrt{-g}) (\delta S / \delta g^{\mu \nu})$. For the vector field, scalar field, and a perfect fluid representing standard matter, the respective energy-momentum tensors are given by:
	\begin{equation}
		T_{\mu \nu}^{(v)} = \frac{1}{4} B^{\alpha\beta} B_{\alpha\beta}g_{\mu\nu} + B^\alpha_\nu B_{\mu\alpha} + \kappa A_\alpha J^\alpha g_{\mu \nu} + 2 \kappa J_\mu A_\nu, 
		\label{eq:T_v}
	\end{equation} 
	\begin{align} 
		T_{\mu \nu}^{(s)}= & 2 \gamma \partial_{\mu} \phi^* \partial_{\nu}\phi + 2 \gamma \omega^2 G A_\mu A_\nu \\
		-&\left(\partial^{\alpha} \phi^* \partial_{\alpha}\phi + \gamma \omega^2 G A^\alpha A_\alpha + \gamma V(G) \right) g_{\mu \nu}\nonumber,
		\label{eq:T_s}
	\end{align} 
	\begin{equation} 
		T_{\mu \nu}^{(M)} = (\rho + p)u_\mu u_\nu + p g_{\mu\nu}.
		\label{eq:T_M}
	\end{equation} 
	
	To determine the evolution of the scale factor $a(t)$, we now derive the generalized Friedmann equations. We construct the Lagrangian density $\mathcal{L}$ corresponding to the action \eqref{eq:s} and apply the Euler-Lagrange equations. Given the symmetries of the FLRW metric, the vector field must be purely temporal, $A^\mu = (A^0(t), 0, 0, 0)$, to respect cosmological isotropy.

The resulting dynamical equations from the Lagrangian and Hamiltonian formulations are, respectively:
\begin{widetext}
	\begin{equation} 
		H^2 + 2\frac{\ddot{a}}{a} + \left(\frac{\dot{G}}{G}\right)^2 - \frac{\ddot{G}}{G} - H\frac{\dot{G}}{G} = \frac{4\pi G}{3} \left[ \kappa a (\rho A_0)' + a\gamma \omega^2 (G A_\mu A^\mu)' + a \gamma V(G)' + a \rho' \right],
		\label{eq:lagran_a}
	\end{equation} 
	
	\begin{equation} 
		H^2 - H \frac{\dot{G}}{G} = \frac{8\pi G}{3} \left[ -\kappa\rho(t) A^0 - \frac{a^5}{2} \sum_{i}(\dot{A}^i)^2 + \gamma \omega^2 G A_\mu A^\mu + \gamma V(G) + \rho(t) \right].
		\label{eq:HHHH}
	\end{equation} 
\end{widetext}
where $H \equiv \dot{a}/a$ is the Hubble parameter, an overdot denotes a derivative with respect to cosmic time $t$, and a prime denotes a derivative with respect to the scale factor $a$. In the limit $\dot{G} = 0$ and $A_\mu = 0$, these equations naturally reduce to the standard Friedmann equations.

Furthermore, the equations of motion for the vector field $A_\mu$ are:
\begin{equation} 
	\ddot{A}^i + 5 H \dot{A}^i + 2\gamma \omega^2 G A^i = 0,
	\label{eq:EulerA_i}
\end{equation} 
\begin{equation} 
	A^0 = - \frac{\kappa \rho}{2 \gamma \omega^2 G}.
	\label{eq:EulerA_0}
\end{equation} 

Equation \eqref{eq:EulerA_i} describes the spatial components $A^i$ as damped harmonic oscillators. Their amplitudes decay rapidly over cosmic time, allowing us to set $A^i \approx 0$. The temporal component $A^0$ is determined entirely by the matter density. Substituting these solutions into Eq. \eqref{eq:HHHH} and defining the coupling parameter $\chi \equiv \kappa^2/(4 \gamma \omega^2)$, we arrive at the modified Friedmann equation:
\begin{equation} 	
	H^2 - H \frac{\dot{G}}{G} = \frac{8\pi G}{3} \left[ \rho + \chi \frac{\rho^2}{G} + \gamma V(G) \right].
	\label{eq:n_Hamil_a}
\end{equation} 

The gravitational coupling $G$ evolves with redshift $z$ as:
\begin{equation} 
	G(z) = \frac{1}{b \beta} \left( \beta - 1 - z + \sqrt{(1 + z - \beta)^2 - m^2 \beta^2 b} \right),
	\label{eq:G_z}
\end{equation} 
where $\beta \equiv a_0/a_c$ is the ratio of the present scale factor to that at the critical epoch $T_c$. The corresponding time derivative is given by:
\begin{equation} 
	\frac{\dot{G}(z)}{G(z)} = H(z) \frac{1+z}{\sqrt{(1+z-\beta)^2-m^2\beta^2 b}}.
	\label{eq:dG_G}
\end{equation} 

Expressing the Friedmann equation \eqref{eq:n_Hamil_a} in terms of redshift $z$ yields the general form of the Hubble parameter:
\begin{equation} 
	H^2(z) = \frac{8 \pi G(z)}{3}\!\left( \rho + \frac{ \chi }{ G(z) }\rho^2 + \gamma V(G)\!\right)\!\left(\!\frac{\mathfrak{B}}{\mathfrak{B} - 1 + z} \right),
	\label{eq:H_General}
\end{equation} 
where
\begin{equation} 
	\mathfrak{B} = \sqrt{(1+z-\beta)^2-m^2\beta^2 b}.
	\label{eq:B_func} 
\end{equation} 
This equation is the primary result of this section and forms the basis for our subsequent observational analysis.
\subsection{Evolution of the Effective Gravitational Constant}
In the GIMOG framework, gravity is an emergent phenomenon. In the very early universe at temperatures far above a critical temperature $T_c$, the scalar field is in a symmetric phase with a vanishing vacuum expectation value, $\langle \phi \bar{\phi} \rangle = 0$. Consequently, the effective gravitational constant is zero, and gravity is inactive.

\begin{figure}[htbp]
	\centering
	\includegraphics[width=1\columnwidth]{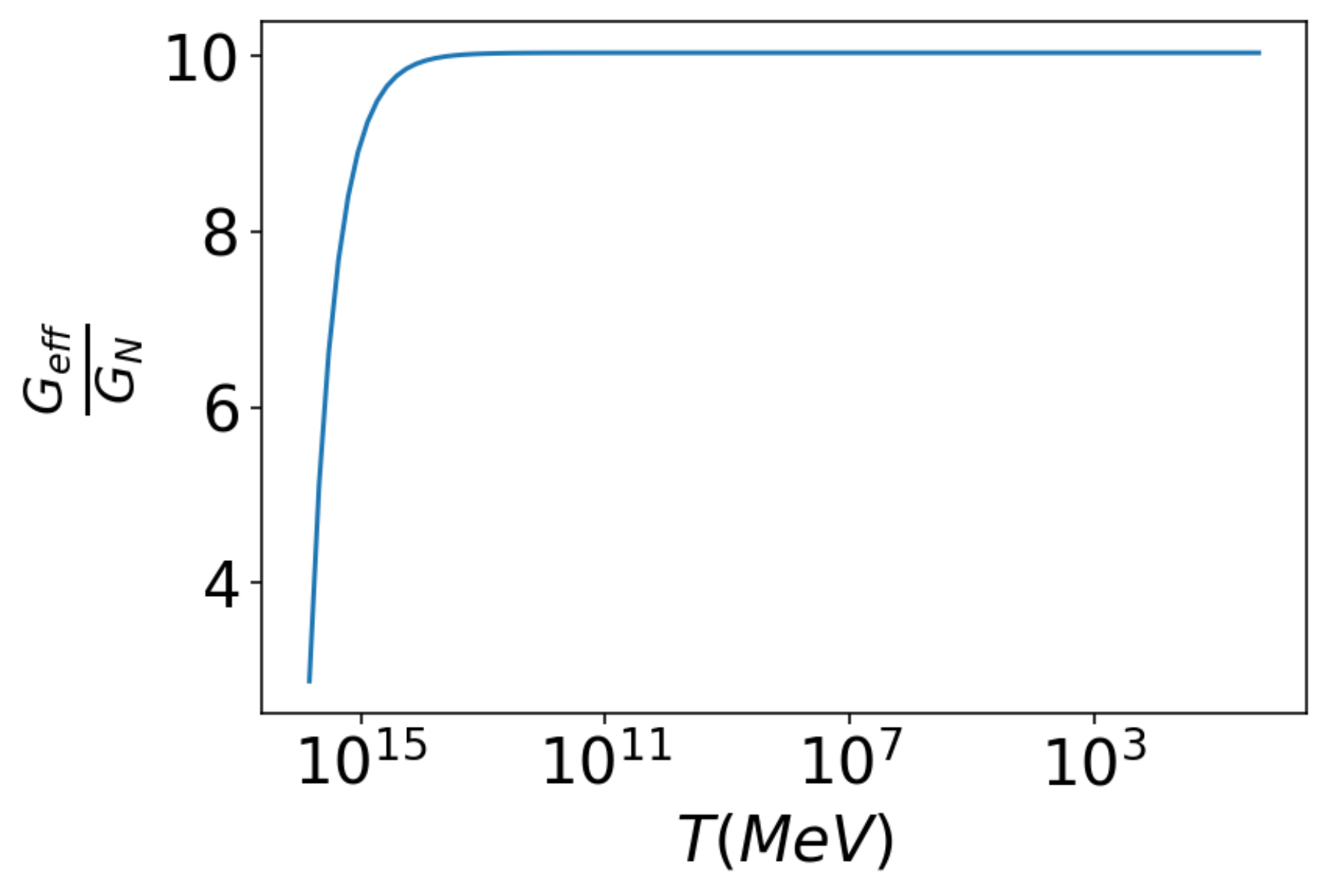}
	\caption{\label{fig:G_a_ac}%
		Evolution of the effective gravitational constant ($G$) as a function of cosmic temperature in the post-inflation epoch, for model parameters $b = 0.2$ and $m^2 = 0.12$. $G$ is scaled to the Newtonian gravitational constant $G_N$.}
\end{figure}

As the universe expands and cools, the temperature drops towards $T_c$. This drives a first-order phase transition, during which the scalar field $\phi$ tunnels from a metastable vacuum to the true minimum of its potential. This transition triggers a brief period of inflation and, crucially, marks the activation of gravity. At the true minimum, the scalar field acquires a non-zero value, which sets the value of the gravitational coupling according to Eq.~\eqref{eq:G_theta}. As shown in Fig.~\ref{fig:G_a_ac}, the value of $G$ increases rapidly from zero to a stable, non-zero value following the phase transition. Subsequently, the evolution of $G$ becomes very slow, ensuring consistency with observational constraints in the radiation- and matter-dominated eras, where $G$ is measured to be nearly constant.
\subsection{The Late-Time Friedmann Equation}\label{M-F-E}
We can derive a simplified form of the Friedmann equation valid for the post-inflationary universe. If we associate $T_c$ with the grand unification scale ($T_{\text{GU}}$), the ratio $\beta = a_0/a_c \approx T_{\text{GU}}/T_0$ is extremely large, on the order of $10^{27}$. For all observationally relevant redshifts ($z \ll \beta$), the final term in Eq.~\eqref{eq:H_General} can be accurately approximated as:
\begin{equation}
	\frac{\mathfrak{B}}{\mathfrak{B} - 1 + z} \approx 1.
\end{equation}

Under this well-justified approximation, the Friedmann equation~\eqref{eq:H_General} simplifies significantly to:
\begin{equation}
	H^2(z) = \frac{8 \pi G}{3}\left( \rho(z) + \frac{\chi}{G} \rho(z)^2 + \gamma V \right),
	\label{H_recent}
\end{equation}
where $G$ is now the nearly constant late-time value. The total energy density $\rho(z)$ includes contributions from matter and radiation, $\rho(z) = \rho_{0m} (1+z)^3 + \rho_{0r} (1+z)^4$.

For the matter-dominated era, we neglect the radiation component. Normalizing the Hubble parameter by its present-day value $H_0$ and using the standard definition of the critical density $\rho_c = 3H_0^2/(8\pi G_N)$, Eq.~\eqref{H_recent} can be written as:
\begin{equation}
	H(z) = H_0 \sqrt{A (1+z)^3 + B (1+z)^6 + C},
	\label{a_lattime}
\end{equation}
where the dimensionless coefficients are defined as:
\begin{align}
	A &= \frac{G}{G_N} \Omega_{0b}, \label{A} \\
	B &= \frac{8 \pi}{3}(\frac{\rho_{0b}}{H_0})^2\chi, \label{B_H} \\
	C &= \frac{8\pi G}{3 H_0^2}\gamma V(G). \label{C}
\end{align}

In this model, we consider only baryonic matter, so $\Omega_{0m}$ is replaced by $\Omega_{0b}$. The parameter $A$ represents the effective matter density, $B$ quantifies the correction from the modified gravity term, and $C$ acts as an effective dark energy component arising from the scalar field potential. For a flat universe, these parameters must satisfy the constraint $A + B + C = 1$. 
\section{Late-Universe Tests}
\label{late universe}
\subsection{Observational Data and Model Fitting}
\label{sec:Sn_method}
In this section, we outline the methodology and observational datasets used to confront the GIMOG model with late-time cosmological probes, beginning with Type Ia Supernovae (SNe Ia) \cite{PantheonData}. 

To compare observational data with our theoretical predictions, we utilize the distance modulus $\mu(z)$, which relates to the luminosity distance $d_L(z)$ as:
\begin{equation}
	\mu(z) = 5 \log_{10} \left( \frac{\mathrm{d}_L(z)}{\text{Mpc}} \right) + 25.
\end{equation}

The luminosity distance is computed from the evolution of the Hubble parameter $H(z)$ via:
\begin{equation}
	\mathrm{d}_L(z) = (1+z) \int_0^z \frac{c \, \mathrm{d}z'}{H(z')}.
\end{equation}

Our primary dataset is the Pantheon+ Supernova sample, a comprehensive compilation consisting of 18 distinct datasets and 1,701 light curves from confirmed SNe Ia. Spanning a redshift range of $0.001 < z < 2.3$ \cite{scolnic2022pantheon+}, this sample provides a robust foundation for constraining cosmological expansion dynamics.

The residuals $\Delta\mu_i$, representing the difference between the observed distance modulus $\mu_{\text{obs},i}$ and the theoretical prediction from the GIMOG model $\mu_{\text{GIMOG},i}$, are given by:
\begin{equation}
	\Delta\mu_i = \mu_{\text{obs},i} - \mu_{\text{GIMOG},i}.
\end{equation}

To rigorously account for both statistical and systematic uncertainties in the dataset, we incorporate the full covariance matrix $C$. The goodness of fit is evaluated using the $\chi^2$ statistic:
\begin{equation}
	\chi^2 = \sum_{i,j} \Delta\mu_i \, (C^{-1})_{ij} \, \Delta\mu_j.
\end{equation}

We determine the best-fit parameters of the GIMOG model by minimizing this $\chi^2$ function. The optimization is performed utilizing Markov Chain Monte Carlo (MCMC) techniques, which efficiently explore the multi-dimensional parameter space and provide robust uncertainty estimates for the derived parameters.

\begin{figure}[htbp]
	\centering
	\includegraphics[width=0.9\columnwidth]{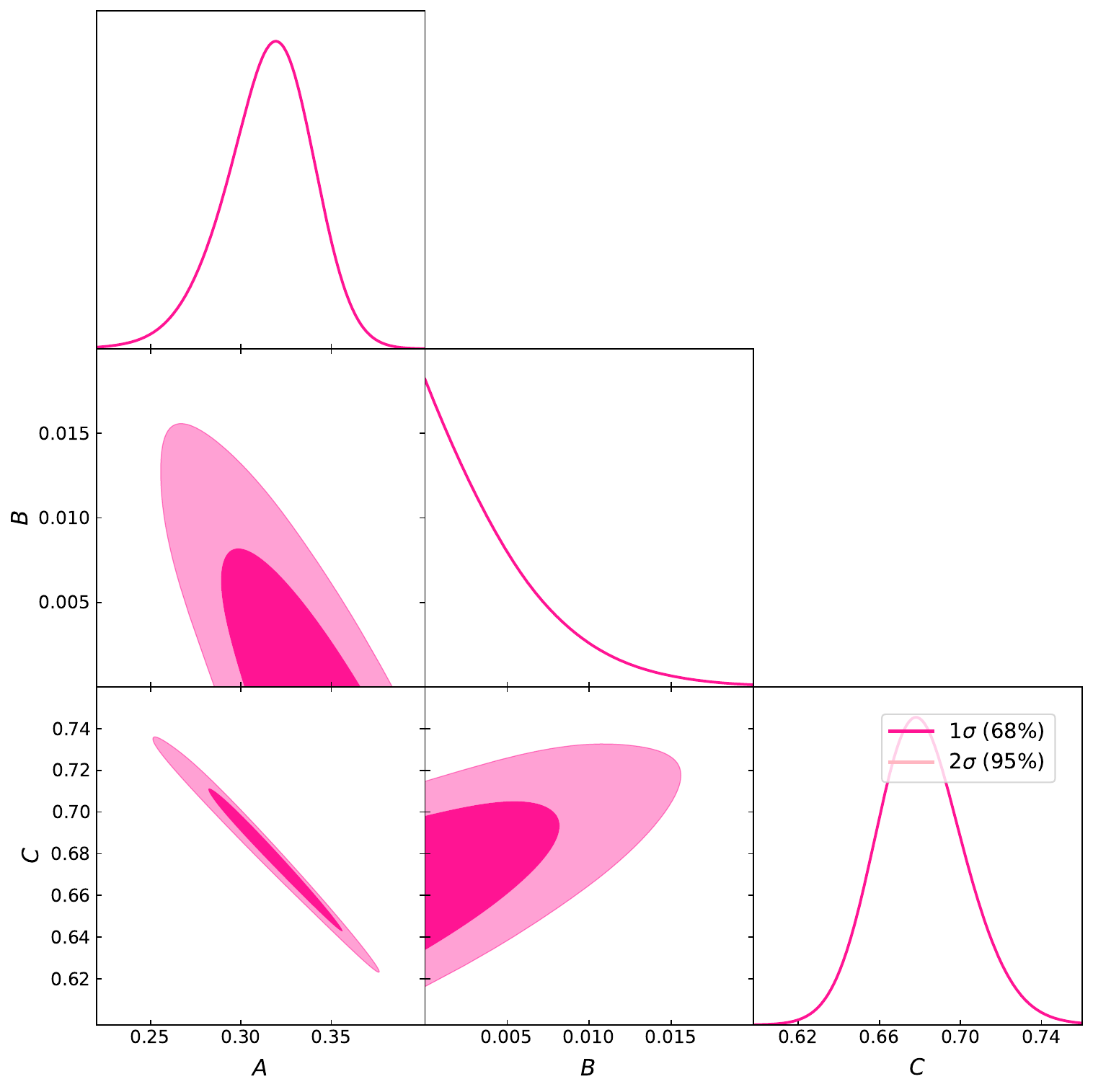}
	\caption{\label{fig:MCMC_SN}%
		The $1\sigma$ ($68\%$) and $2\sigma$ ($95\%$) marginalized confidence contours and 1D probability distributions for the effective GIMOG density parameters: $A$ (matter contribution), $B$ (higher-order gravitational correction), and $C$ (effective cosmological constant), reconstructed from the Pantheon+ supernova dataset. 
		The 1D marginalized distribution of $B$ exhibits a well-behaved semi-Gaussian profile peaked near zero, effectively imposing a stringent upper bound on high-energy modifications in the late-time Universe.}
\end{figure}

\begin{figure}[htbp]
	\centering
	\includegraphics[width=1\columnwidth]{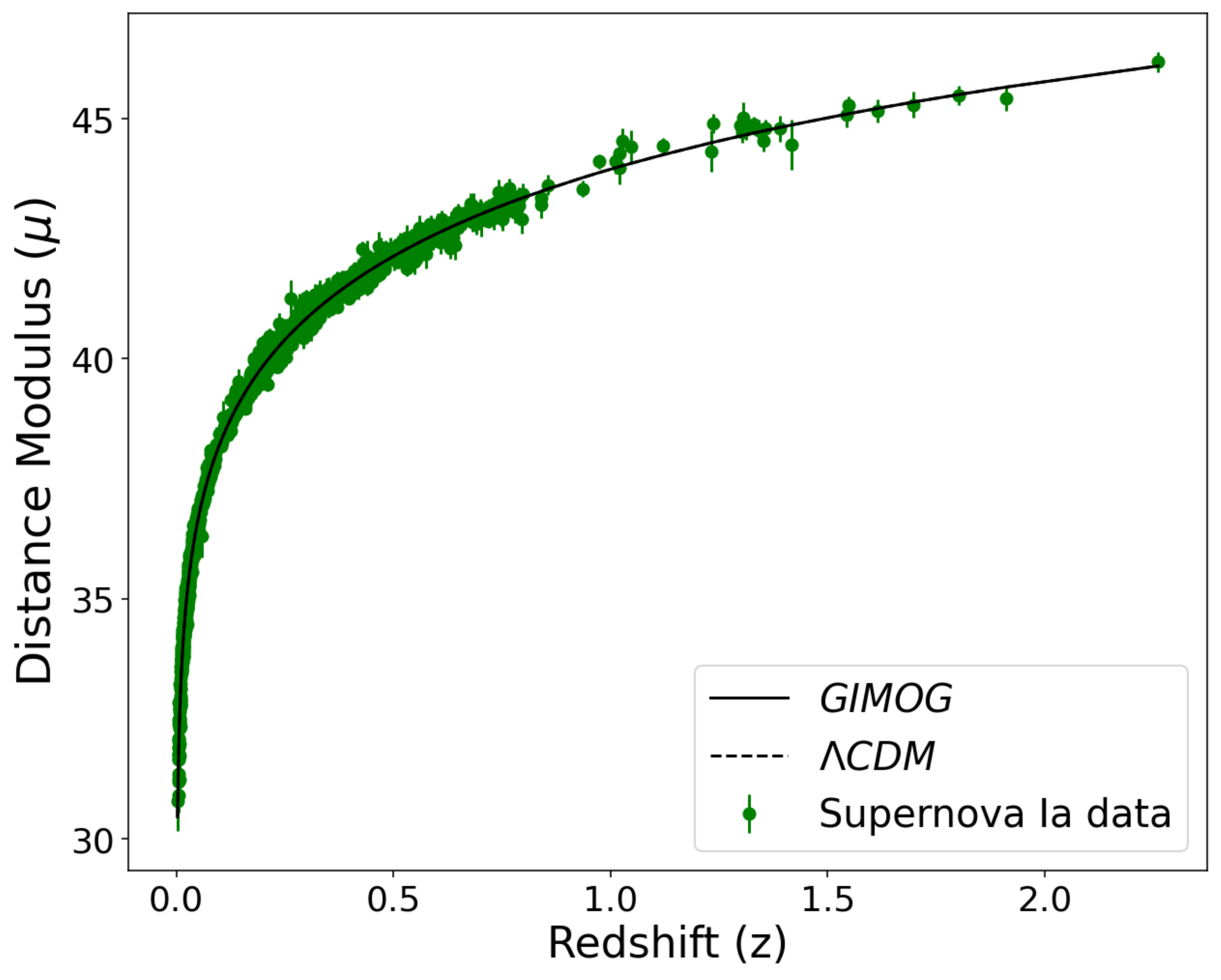}
	\caption{\label{fig:mu}%
		The distance modulus $\mu(z)$ as a function of redshift $z$ for the Pantheon+ Type Ia Supernova dataset. The points represent the observational data with error bars. The dashed line corresponds to the $\Lambda\text{CDM}$ model fit, while the solid line represents the best-fit curve from our GIMOG model.}
\end{figure}

The results of this MCMC fitting process are illustrated in Fig.~\ref{fig:MCMC_SN}, detailing the posterior distributions, and Fig.~\ref{fig:mu}, which compares the best-fit GIMOG distance modulus with both the observational data and the standard $\Lambda$CDM model. Notably, the theoretical curves for GIMOG and $\Lambda$CDM are nearly indistinguishable in this redshift regime. 

The parameter $A$ closely mirrors the standard matter-density parameter, while $C$ behaves as an effective dark-energy-like contribution associated with the scalar-field potential. The extremely small value of $B$ indicates that the higher-order modified-gravity correction to the expansion rate is negligible at late times.

The best-fit parameters derived from the analysis are summarized in Table~\ref{tab:ta1}. Using the best-fit value $A=0.3150$ and adopting a fiducial baryon density $\Omega_{0b}=0.048$, the relation $A=(G/G_N)\Omega_{0b}$ yields $G/G_N \simeq 6.6$. It is important to emphasize, however, that the Pantheon+ sample directly constrains the effective matter-like coefficient $A$, rather than $G/G_N$ itself. The inference $G/G_N \simeq 6.6$ follows only under the baryon-only assumption of the present model and the use of an external prior on $\Omega_{0b}$. Within this interpretation, the late-time expansion data favor an effective gravitational coupling stronger than the Newtonian value, qualitatively similar to the large-scale behavior of MOG, where $G/G_N=1+\alpha$ with $\alpha=8.89\pm0.34$ from galactic rotation-curve fits \cite{moffat2013mog}. It should be stressed that this large value of $G/G_N$ refers to an effective large-scale gravitational coupling and should not be interpreted as a uniform rescaling of Newton's constant in local gravitational tests. In the weak-field formulation of MOG, the enhanced long-range attraction is accompanied by a repulsive Yukawa contribution from the vector field, allowing the Newtonian limit to be recovered at sufficiently small scales \cite{moffat2013mog}.

\begin{table}[htbp]
	\centering
	\caption{Best-fit parameters and their corresponding $68\%$ confidence limits for the GIMOG model obtained from the Pantheon+ dataset analysis.}
	\label{tab:ta1}
	\begin{tabular*}{\columnwidth}{@{\extracolsep{\fill}}lcc}
		\hline
		\hline
		Parameter & Best-fit Value & Uncertainty ($1\sigma$) \\
		\hline
		$A$       & $0.3150$ & $^{+0.0269}_{-0.0203}$ \\
		$B$       & $ < 0.0044$  &   \\
		$C$       & $0.6806$  & $^{+0.0198}_{-0.0236}$ \\
		\hline
		\hline
	\end{tabular*}
\end{table}

Collectively, these parameters demonstrate that the GIMOG model successfully replicates standard $\Lambda$CDM-like behavior at the present epoch.
\subsection{Comparison with Cosmic Chronometers}
\label{sec:comparison}
The cosmic chronometer (CC) method provides a direct, model-independent estimate of the Hubble parameter by measuring the expansion rate $dz/dt$ \cite{jimenez2002constraining}. This is achieved by determining the age difference, $\Delta t$, between pairs of passively evolving galaxies that formed at the same epoch but are separated by a small redshift interval, $\Delta z$. By approximating $dz/dt \approx \Delta z/\Delta t$, the Hubble parameter is evaluated as:
\begin{equation}
	H(z) = -\frac{1}{1+z} \frac{\mathrm{d}z}{\mathrm{d}t}.
\end{equation}

In contrast to the high-precision Pantheon+ sample, the CC dataset is significantly sparser, containing significantly fewer independent measurements over a comparable redshift range. Given this disparity in statistical weight and the absence of a common covariance structure between the two probes, performing a simultaneous MCMC fit would disproportionately bias the parameter space toward the SNe Ia constraints while underutilizing the CC data's primary strength: their model independence.

Consequently, we take the best-fit parameters derived exclusively from the Pantheon+ sample (Table~\ref{tab:ta1}) as fixed inputs and compute the theoretical $H(z)$ trajectory from the GIMOG model. We then compare this trajectory directly against the CC1 \cite{simon2005constraints} and CC2 \cite{moresco20166, negrelli2020testing} measurements without any further optimization. This approach allows us to assess whether a model calibrated solely by distance modulus data can accurately predict the expansion history $H(z)$ over the redshift range where both datasets overlap.

\begin{figure}[htbp]
	\centering
	\includegraphics[width=1\columnwidth]{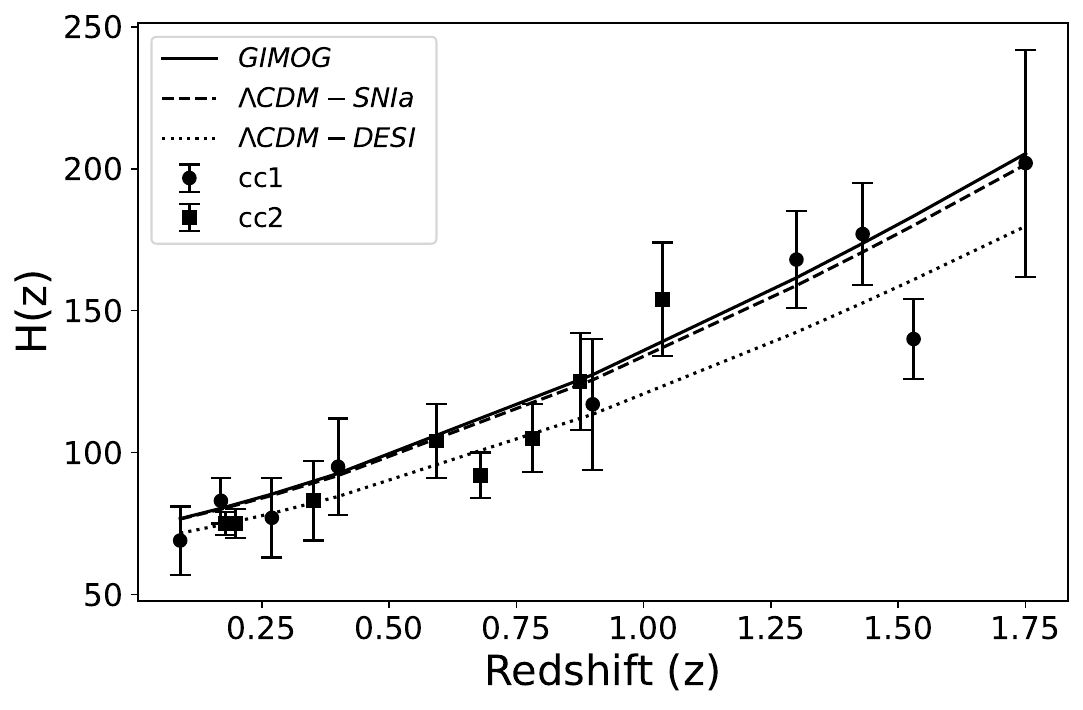}
	\caption{\label{fig:CC}%
		Evolution of the Hubble parameter $H(z)$ as a function of redshift $z$. The dashed line corresponds to the $\Lambda$CDM model constrained by SN Ia data \cite{riess2022comprehensive}, and the dotted line shows the $\Lambda$CDM model based on DESI constraints \cite{adame2025desi}. The solid line represents the GIMOG model prediction using the best-fit parameters from Table~\ref{tab:ta1} (derived solely from SNe Ia). Additionally, independent $H(z)$ measurements from the CC1 \cite{simon2005constraints} and CC2 \cite{moresco20166, negrelli2020testing} datasets are shown as discrete data points with error bars.}
\end{figure}

In Fig.~\ref{fig:CC}, we present this comparison alongside two $\Lambda$CDM reference models: one constrained by DESI full-shape data ($\Omega_{0m}=0.296\pm0.010$, $H_0=68.63\pm0.79$ km/s/Mpc; dotted line) and the other by the Pantheon+ sample ($\Omega_{0m}=0.334\pm0.018$, $H_0=73.04\pm1.04$ km/s/Mpc; dashed line). The solid line represents the GIMOG prediction using the parameters derived from our MCMC analysis (Table~\ref{tab:ta1}).

Thus, the GIMOG prediction exhibits good consistency with the independent cosmic chronometer measurements across the overlapping redshift range
\section{Primordial Universe Tests}
\label{sec:constrain}
\subsection{Cosmic Expansion Before and During the Symmetry Breaking}
In the GIMOG framework, the early Universe undergoes a distinct phase of expansion prior to the dynamical activation of gravity and the onset of inflation. At temperatures exceeding the critical point $T_c$, the scalar field $\phi$ resides in a symmetric vacuum state, characterized by a vanishing vacuum expectation value, $\langle \phi \bar{\phi} \rangle = 0$. Consequently, the effective gravitational constant $G$ is zero, rendering gravitational interactions effectively absent during this primordial epoch.

Despite this absence of gravity, cosmological expansion is not halted. The modified Friedmann equation in this model incorporates a non-linear term of the form $\chi \rho^2$. This term, interpreted as a high-energy correction originating from non-linear matter contributions, drives a non-inflationary but monotonic expansion in the early Universe. Assuming a radiation-dominated equation of state, this mechanism yields a scale factor evolution scaling approximately as $a(t) \propto t^{1/4}$, which proceeds more slowly than the standard $t^{1/2}$ expansion.

As the temperature drops below $T_c$, the scalar potential undergoes a first-order phase transition. The field tunnels from the metastable symmetric vacuum to a lower-energy true vacuum, liberating vacuum energy. This process triggers a phase of rapid accelerated expansion (inflation), which is now governed by a nearly constant scalar potential $V(G)$ and a non-zero, dynamically generated $G$ field as $G = \langle \phi \bar{\phi} \rangle$. The dynamics therefore describe a transition from a pre-inflationary slow expansion—driven by non-gravitational corrections—to an inflationary phase exhibiting effective de Sitter-like behavior \cite{rouhani2024mog}.

This two-stage expansion scenario, consisting of pre-gravitational growth followed by symmetry-breaking-driven inflation, provides a natural mechanism for the emergence of gravity. Furthermore, it inherently connects the initial conditions of inflation to the underlying scalar field dynamics. This framework offers a consistent cosmological timeline that begins in a high-temperature, gravity-free regime and smoothly transitions into a conventional radiation-dominated Universe following reheating.

Immediately after inflation concludes, the vacuum energy is converted into radiation during the reheating phase, populating the Universe with Standard Model particles. From this point onward, gravity enters into the cosmological field equations, and the expansion of the Universe proceeds under the influence of the established gravitational constant, $G$.
\subsection{Big Bang Nucleosynthesis}
Big Bang Nucleosynthesis (BBN) stands as a cornerstone prediction of the standard cosmological model, offering a compelling explanation for the primordial origin of light elements. Occurring within the first few minutes after the Big Bang, BBN describes the epoch when the hot, dense Universe cooled sufficiently to permit the formation of stable atomic nuclei beyond free protons. Crucially, during the earliest, ultra-hot phase, baryonic matter existed predominantly as free protons and neutrons. These nucleons were maintained in thermal equilibrium via weak interaction processes:
\begin{gather*}
	\nu_e + n \rightleftharpoons p + e^-, \\
	e^+ + n \rightleftharpoons p + \bar{\nu}_e, \\
	n \rightleftharpoons p + e^- + \bar{\nu}_e.
\end{gather*}

The neutron-to-proton ratio dictates the initial conditions for BBN and fundamentally determines the primordial abundance of $^4\text{He}$ (for detailed calculations, see \cite{padmanabhan1993structure}).

In kinetic equilibrium, the number density $n_A$ of a highly nonrelativistic nucleus follows:
\begin{equation}
	n_A = g_A \left( \frac{m_A T}{2 \pi} \right)^{3/2} \exp\left( \frac{\mu_A - m_A}{T} \right).    
\end{equation}
In the above relation, $\mu_A$ represents the chemical potential of the nucleus $A(Z)$. Under thermodynamic equilibrium conditions—when the rate of nuclear reactions vastly exceeds the cosmic expansion rate—chemical equilibrium is maintained and the following relationship holds:
\begin{equation}
	\mu_A = Z \mu_p + (A - Z) \mu_n.
\end{equation} 
Applying these relations to protons and neutrons, and approximating $m_n \simeq m_p \simeq m_A/A \equiv m_N$ in the pre-exponential factors, the abundance of species $A(Z)$ is given by the Saha equation:
\begin{equation}
	n_A =g_A A^{3/2 - A} \left( \frac{2 \pi}{m_N T} \right)^{3(A-1)/2}\!\!\!
	n_p^{Z} n_n^{A-Z} \exp\left( \frac{B_A}{T} \right), 
\end{equation} 
where $B_A$ is the binding energy and $g_A$ is the statistical factor. Using the mass fraction,
\begin{equation}
	X_A \equiv \frac{n_A A}{n_b}, 
\end{equation}   
the equilibrium abundance in nuclear statistical equilibrium (NSE) takes the form:
\begin{eqnarray}
	X_A =&& g_A \zeta(3)^{A-1} \pi^{(1-A)/2} 2^{(3A-5)/2} A^{5/2}\\
	&&\times \left( \frac{T}{m_b} \right)^{3(A-1)/2} \eta^{A-1} X_p^Z X_n^{A-Z} \exp\left( \frac{B_A}{T} \right),\nonumber
	\label{eq:Saha}
\end{eqnarray}
with $\eta \equiv n_b/n_\gamma$ denoting the baryon-to-photon ratio.

When the rates for these interactions are fast compared to the expansion rate $H$, chemical equilibrium is maintained:
\begin{equation}
\mu_n + \mu_\nu = \mu_p + \mu_e,
\end{equation} 
from which it follows that the neutron-to-proton ratio in chemical equilibrium is:
\begin{equation}
\frac{n}{p} \equiv \frac{n_n}{n_p} = \frac{X_n}{X_p} = \exp\left[ -\frac{Q}{T} + \frac{(\mu_e - \mu_\nu)}{T} \right],
\end{equation} 
where $Q \equiv m_n - m_p = 1.293 \, \text{MeV}$. Under the standard assumption of vanishing lepton chemical potentials, the equilibrium neutron-to-proton ratio simplifies to:
\begin{equation}
\left.\frac{n}{p}\right|_{\text{eq}} = \exp\left(-\frac{Q}{T}\right).
\label{eq:n_p}
\end{equation}  

When the interaction rate becomes comparable to the Hubble expansion rate ($\Gamma \sim H$), weak interactions cease to maintain equilibrium, thereby fixing the neutron-to-proton ratio \cite{kolb2018early}. 

The weak interaction rates governing $\beta$-processes ($n \rightleftharpoons p$) are determined by thermally averaging the cross-sections over the momentum distributions of relativistic species. The rate per target nucleon is obtained through the phase-space integral. Defining $\mathcal{G} = G_F^2 (1 + 3g_A^2)/(2\pi^3)$, these reaction rates are:
\begin{equation}
\Gamma_{\nu_e n} = \mathcal{G} \int_0^\infty \!\mathrm{d}E_\nu E_\nu^2 E_e \sqrt{E_e^2 - m_e^2} \, f_{\nu_e}(E_\nu) \left[1 - f_{e^-}(E_e)\right],
\end{equation} 
where $E_e = E_\nu + Q$.
\begin{equation}
\Gamma_{e^+ n} =\!\!\mathcal{G} \int_{Q + m_e}^\infty\!\!\mathrm{d}E_\nu E_\nu^2 E_e \sqrt{E_e^2 - m_e^2} \,f_{e^+}(E_e) \left[1 - \!f_{\bar{\nu}_e}(E_\nu)\right] ,
\end{equation} 
where $E_e = E_\nu - Q$.
\begin{equation}
\Gamma_{e^- p} = \mathcal{G} \int_{m_e}^\infty \mathrm{d}E_e E_\nu^2 E_e \sqrt{E_e^2 - m_e^2} \, f_{e^-}(E_e) \left[1 - f_{\nu_e}(E_\nu)\right],
\end{equation} 
where $E_\nu = E_e - Q$.
\begin{equation}
\Gamma_{\bar{\nu}_e p} = \!\mathcal{G}\! \int_{Q + m_e}^\infty \!\mathrm{d}E_\nu E_\nu^2 E_e \sqrt{E_e^2 - m_e^2} \, f_{\bar{\nu}_e}(E_\nu) \left[1 - \!f_{e^+}(E_e)\right],
\end{equation} 
where $E_e = E_\nu - Q$.
\begin{equation}
\Gamma_{\bar{\nu}_e e^- p} = \mathcal{G} \int_0^{Q} \mathrm{d}E_\nu E_\nu^2 E_e \sqrt{E_e^2 - m_e^2} \, f_{\bar{\nu}_e}(E_\nu) f_{e^-}(E_e),
\end{equation} 
where $E_e = Q - E_\nu$.

The rate of interaction relates to the interaction timescale by:
\begin{equation}
\Gamma_{n} = \frac{1}{\tau_n}.
\end{equation} 

\begin{figure}[htbp]
\centering
\includegraphics[width=1\columnwidth]{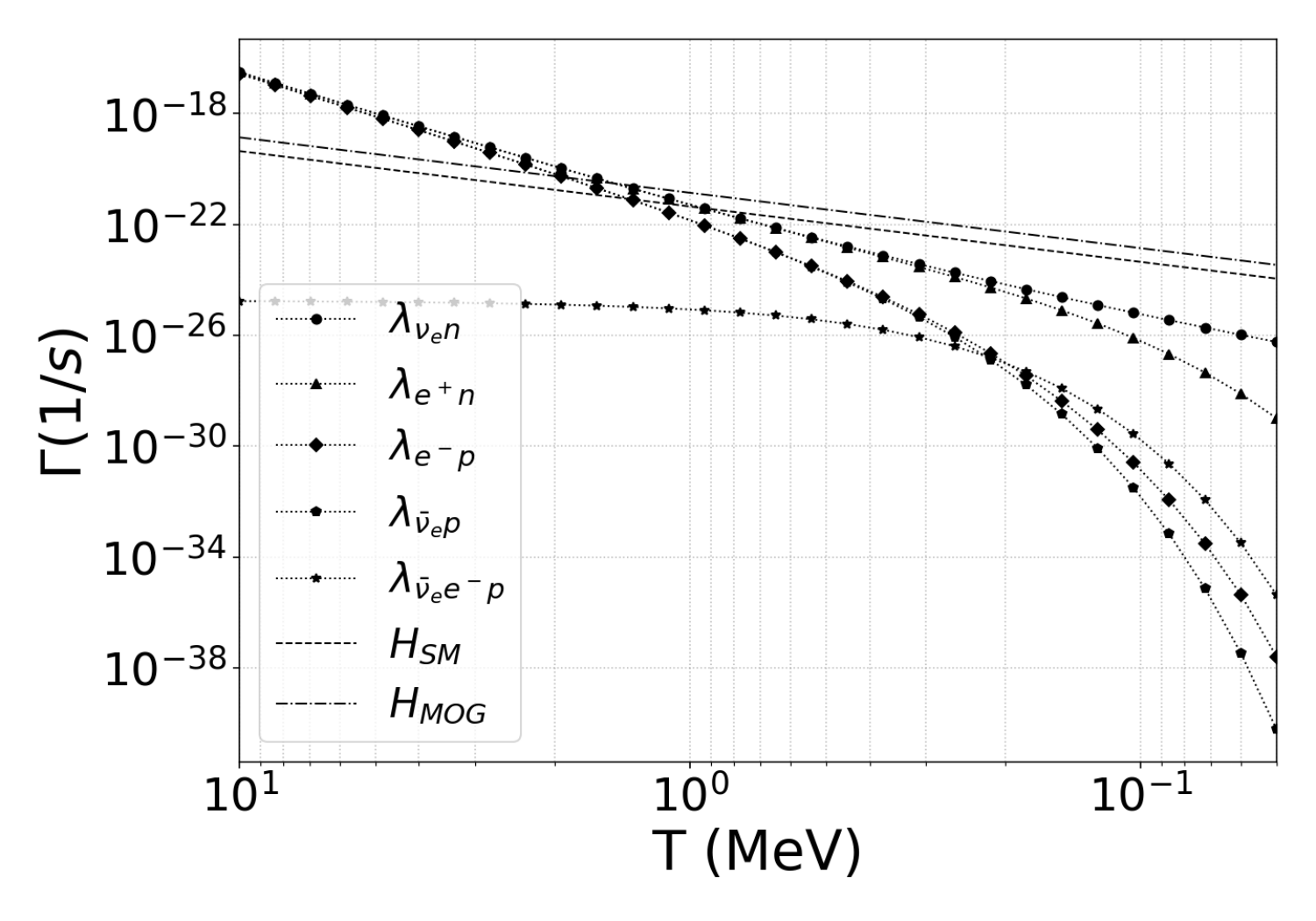}
\caption{Reaction rates and cosmic expansion rates plotted as a function of temperature $T$. The curves illustrate the weak interaction rates for individual $\beta$-processes ($\Gamma_{\nu_e n}$, $\Gamma_{e^+ n}$, $\Gamma_{e^- p}$, $\Gamma_{\bar{\nu}_e p}$, and $\Gamma_{\bar{\nu}_e e^- p}$) alongside the total expansion rate. The Hubble parameter is shown for both the standard cosmological model ($H_{\text{SM}}$) and the modified gravity framework ($H_{\text{GIMOG}}$).}
\label{fig:Gamma}
\end{figure}

Here, we utilize the Fermi-Dirac distribution function for a fermionic particle, given by:
\begin{equation}
f(E; T, \mu) = \frac{1}{\exp\left(\frac{E-\mu}{T}\right) + 1},
\end{equation} 
where $\mu$ is the chemical potential and $T$ is the temperature of the cosmic plasma. Fig.~\ref{fig:Gamma} plots the reaction rates as a function of cosmic temperature around the epoch of freeze-out.

\subsubsection{BBN in MOG}
We first evaluate the abundance of light elements resulting from BBN within the standard MOG framework. The expansion rate of the Universe is given by:
\begin{equation}
H^2 = \frac{8\pi G_N}{3} \rho_r,
\end{equation}
where $\rho_r$ denotes the radiation energy density:
\begin{equation}
\rho_r = \frac{\pi^2}{30} g_* T^4.
\end{equation}

In MOG, the gravitational constant $G$ is modified as \cite{moffat2009fundamental}:
\begin{equation}
G = (1 + \alpha) G_N.
\end{equation}

Based on fits to the dynamics of galaxies and galaxy clusters \cite{moffat2013mog}, the preferred value of $\alpha$ is:
\begin{equation}
\alpha = 8.89 \pm 0.34.
\end{equation}

To determine the freeze-out abundance, we solve the Boltzmann equation \cite{baumann2022cosmology, dodelson2024modern}:
\begin{equation}
\frac{\mathrm{d}X_n}{\mathrm{d}t} = -\Gamma_n \left[ X_n - (1 - X_n) e^{-Q/T} \right]. 
\end{equation}

Introducing the dimensionless variable $x = Q/T$ and defining the Hubble parameter parameters as:
\begin{equation}
H_1 = \frac{2\pi Q^2}{3} \sqrt{\frac{\pi g_* G_N (1+ \alpha)}{5}},
\end{equation}
where $H(x) = H_1/x^2$, we obtain the modified differential equation:
\begin{equation}
\frac{\mathrm{d}X_n}{\mathrm{d}x} = \frac{\Gamma_n (x)}{H_1} x \left[ e^{-x} - X_n \left(1 + e^{-x}\right) \right].
\end{equation}   

\begin{figure}[htbp]
\centering
\includegraphics[width=1\columnwidth]{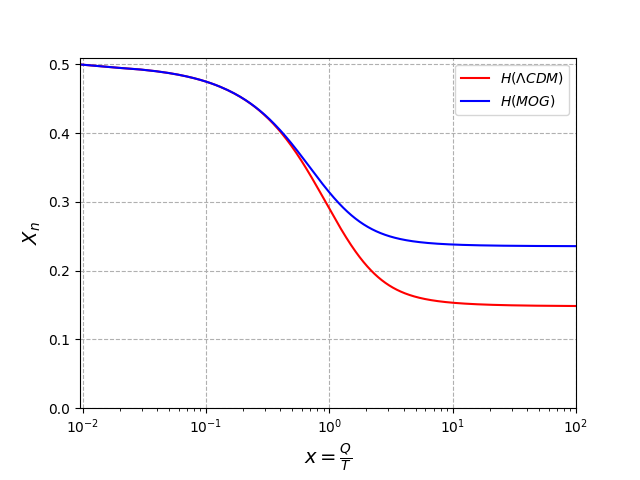}
\caption{The neutron mass fraction as a function of the dimensionless variable $x = Q/T$ for both the standard cosmological model (SM) and modified gravity (MOG).}
\label{fig:X_n}
\end{figure}

Numerical solution of this equation yields the neutron-to-proton freeze-out ratio. Fig.~\ref{fig:X_n} displays the neutron mass fraction as a function of $x$ for both the standard cosmological model (SM) and modified gravity (MOG). We observe that the asymptotic mass fraction of neutrons is $X_n^\infty \simeq 0.15$ in the SM, whereas it rises to $X_n^\infty \simeq 0.24$ in MOG.

As shown in Fig.~\ref{fig:Gamma}, the enhanced expansion rate in MOG causes weak interactions to depart from equilibrium at a higher temperature than in the standard model. Consequently, neutron-proton decoupling occurs earlier, resulting in a higher neutron-to-proton ratio at freeze-out. Thus, the neutron-to-proton freeze-out occurs at $T \simeq 0.8 \, \text{MeV}$ in the SM and at $T \simeq 1.2 \, \text{MeV}$ in MOG (see Fig.~\ref{fig:Gamma}). The resulting neutron-to-proton ratio at freeze-out in MOG is therefore:
\begin{equation}
\frac{n}{p} = \exp\left(-\frac{Q}{T_f}\right) = 0.32.    
\end{equation}

Following the weak-interaction freeze-out, the neutron-to-proton ratio ($n/p$) exhibits a gradual departure from its equilibrium value due to residual weak processes, primarily free neutron decay. This deviation becomes significant by the onset of nucleosynthesis.

Although nuclear binding energies per nucleon are substantial ($\mathcal{O}(1-10)\,\text{MeV}$), thermodynamic equilibrium favors light nuclei (e.g., deuterium, helium) only below temperatures of $\mathcal{O}(0.1-1)\,\text{MeV}$. This delay arises from the high entropy of the Universe, quantified by the small baryon-to-photon ratio ($\eta$). While energy considerations favor bound nuclei at temperatures below a few MeV, entropy dominance preferentially maintains free nucleons until significantly lower temperatures are reached.

The characteristic formation temperature ($T_{\text{NUC}}$) for a nuclear species $A$ is derived from Eq.~\eqref{eq:Saha} by applying the criterion $X_A \sim 1$, signifying a sudden phase transition from free nucleons to nuclear dominance. This condition emerges from the competition between the binding energy ($B_A/T$) and entropy contributions ($\propto \ln \eta^{-1}$), triggering rapid nuclear synthesis when $\exp(B_A/T)$ overcomes the Boltzmann suppression. This approximation yields:
\begin{equation}
T_{\text{NUC}} \approx \frac{B_A/(A - 1)}{\ln(\eta^{-1}) + \frac{3}{2} \ln(m_N/T)}.
\label{eq:T_NUC_approx}
\end{equation}

For deuterium ($A=2$), the characteristic formation temperature is $T_{\text{NUC}} \approx 0.07\,\text{MeV}$. The corresponding timescale for deuterium synthesis is derived from the Hubble parameter evolution $H = 1/(2t)$ in a radiation-dominated Universe, yielding:
\begin{equation}
t(s) = \frac{1}{\sqrt{g_* (1+\alpha)}}\left(\frac{1.5}{T_{\text{MeV}}}\right)^2.
\end{equation}

By the nucleosynthesis epoch, $g_*$ reduces to its present value ($g_* = 3.36$) due to electron-positron annihilation and the subsequent entropy transfer to photons. Substituting $T_{\text{NUC}}$ gives:
\begin{equation*}
t_{\text{NUC}} \simeq 79.22\,\text{s}.
\end{equation*}

The neutron mass fraction at the time of helium-4 production, accounting for neutron decay since freeze-out, follows:
\begin{equation}
X_n^{\text{NUC}} = X_n^{\infty} \exp\left(-\frac{t_{\text{NUC}}}{\tau_n}\right),
\label{eq:X_NUC}
\end{equation}
where $\tau_n$ is the neutron lifetime, and $X_n^{\infty}$ denotes the freeze-out abundance.

The standard calculation of the primordial helium-4 mass fraction ($Y_p$) assumes the near-complete incorporation of all available neutrons into $^4\text{He}$ nuclei. This approximation is well-justified due to the absence of stable mass-5 or mass-8 nuclear bottlenecks, the exponential suppression of deuterium photodissociation ($\exp(-B_A/T)$) below $T \simeq 0.07\,\text{MeV}$, and the rapid reaction kinetics ($H^2 \rightarrow H^3 \rightarrow \text{He}^3 \rightarrow \text{He}^4$) that outpace the Hubble expansion by several orders of magnitude. Under these conditions:
\begin{equation}
Y_p = \frac{4 n_{\text{He4}}}{n_n + n_p} = \frac{2n/p}{1+n/p} = 2 X_n^{\text{NUC}}.
\label{eq:Y_p}
\end{equation}

In MOG cosmology, this yields:
\begin{equation*}
Y_p \simeq 0.43.   
\end{equation*}

This theoretical prediction is in severe tension with the observationally determined primordial helium abundance, $Y_p^{\rm obs}=0.2451\pm0.0026$, inferred from spectroscopic analyses of the metal-poor HII region NGC 346 \cite{valerdi2019determination}.
\subsubsection{BBN in GIMOG}
\label{sec:He_method}
We now derive the abundance of primordial $^4\text{He}$ within the Gauge Invariant Modified Gravity (GIMOG) framework. The expansion rate of the Universe is given by Eq.~\eqref{H_recent}, which governs the thermal history of the early Universe and directly impacts BBN. The modified expansion rate alters the freeze-out temperature of weak interactions and the neutron-to-proton ratio, thereby influencing the primordial $^4\text{He}$ yield. In the radiation-dominated epoch, the expansion rate of the Universe is defined by:
\begin{equation}
H^2(T) = \frac{8 \pi^3 G g_*}{90} T^4 + \frac{8 \pi 5 g_*^2 \chi}{2700} T^8.
\label{eq:H_T}
\end{equation} 

Since the scale factor evolves inversely with temperature ($a \propto 1/T$), taking the total logarithmic derivative of this relation yields the kinematic identity:
\begin{equation}
\frac{\mathrm{d}a}{a} = -\frac{\mathrm{d}T}{T}.
\label{eq:app_kinematic}
\end{equation}
By expressing the Hubble parameter through this relation as $H = -(1/T)(\mathrm{d}T/\mathrm{d}t)$, Eq.~\eqref{eq:H_T} can be recast into a first-order differential equation for the temperature evolution, namely $\mathrm{d}t = -\mathrm{d}T / (T^3 \sqrt{\tilde{A} + \tilde{B} T^4})$. Integrating this relation from the initial singularity ($t \to 0$ as $T \to \infty$) to a given temperature $T$ and employing the change of variable $u = 1/T^2$, we obtain the exact analytical expression for the cosmic time-temperature profile:
\begin{equation}
t(T) = \frac{1}{2\tilde{A}} \left( \sqrt{\frac{\tilde{A}}{T^4} + \tilde{B}} - \sqrt{\tilde{B}} \right),
\label{eq:final_t_T}
\end{equation}
where the constant coefficients are defined as:
\begin{equation}
\tilde{A} = \frac{8 \pi^3 G g_*}{90}, \qquad \tilde{B} = \frac{8\pi^5 g_*^2\chi}{2700}.
\end{equation}

It is worth noting that Eq.~\eqref{eq:final_t_T} consistently recovers the standard cosmological relation in the general relativity limit. By setting the modified gravity parameter to zero, i.e., $\chi \to 0$ (which implies $\tilde{B} = 0$), the expression directly and seamlessly reduces to the well-known standard radiation-dominated time-temperature profile:
\begin{equation}
t(T) = \frac{1}{2\sqrt{\tilde{A}} T^2} = \left( \frac{90}{32\pi^3 G g_*} \right)^{1/2} \frac{1}{T^2}.
\label{eq:standard_t_T}
\end{equation}

The neutron-to-proton ratio ($n/p$) freezes out when the weak interaction rate $\Gamma_{p \to n}$ falls below the Hubble expansion rate $H$. At the freeze-out temperature $T_f$, this condition reads: $\Gamma(T_f) \simeq H(T_f)$. For the $p \to n$ conversion rate, we adopt the Fermi theory approximation \cite{kolb1991early}, $\Gamma_{p \to n} \approx A_w G_F^2 T^5$, where $G_F$ is the Fermi constant and $A_w$ is a dimensionless scaling factor. Balancing the square of this rate with the generalized Hubble parameter leads to a polynomial equation for $T_f$:
\begin{equation}
A_w^2 G_F^4 T_f^6 - \tilde{B}T_f^4 - \tilde{A} \simeq 0.
\label{eq:condition_Tf}
\end{equation}

The underlying parameter space of this generalized gravitational framework is characterized by three independent variables: (i) the normalized effective gravitational constant $G/G_N$, (ii) the parameter $\chi$ quantifying deviations from General Relativity (GR), and (iii) the baryon-to-photon ratio $\eta$. These parameters are rigorously constrained using observational bounds on the primordial Helium-4 mass fraction, $Y_p$.

Operationally, the freeze-out temperature $T_f$ is determined by numerically solving Eq.~\eqref{eq:condition_Tf}, which subsequently fixes the initial neutron-to-proton ratio via standard equilibrium statistics. The onset of deuterium synthesis occurs at a critical temperature $T_{\text{NUC}}$, derived from nuclear binding dynamics via Eq.~\eqref{eq:T_NUC_approx}. The corresponding cosmic timescale, $t_{\text{NUC}} \equiv t(T_{\text{NUC}})$, is then directly obtained by evaluating the analytical time-temperature profile given in Eq.~\eqref{eq:final_t_T} at the nucleosynthesis temperature.

The surviving neutron fraction at the onset of BBN is governed by the standard exponential beta-decay factor, $\exp(-t_{\rm NUC}/\tau_n)$. To constrain the joint parameter space $(G/G_N,\chi,\eta)$, we implement a Bayesian parameter-estimation pipeline based on Markov Chain Monte Carlo (MCMC) sampling. For consistency with the observational baseline adopted in the preceding analysis, the numerical framework is calibrated against the primordial helium abundance
$Y_p^{\rm obs}=0.2451\pm0.0026$,
as inferred from spectroscopic observations of the metal-poor HII region NGC 346 \cite{valerdi2019determination}.
\subsection{Constraining Result}	
\begin{figure}[htbp]
\centering
\includegraphics[width=1\columnwidth]{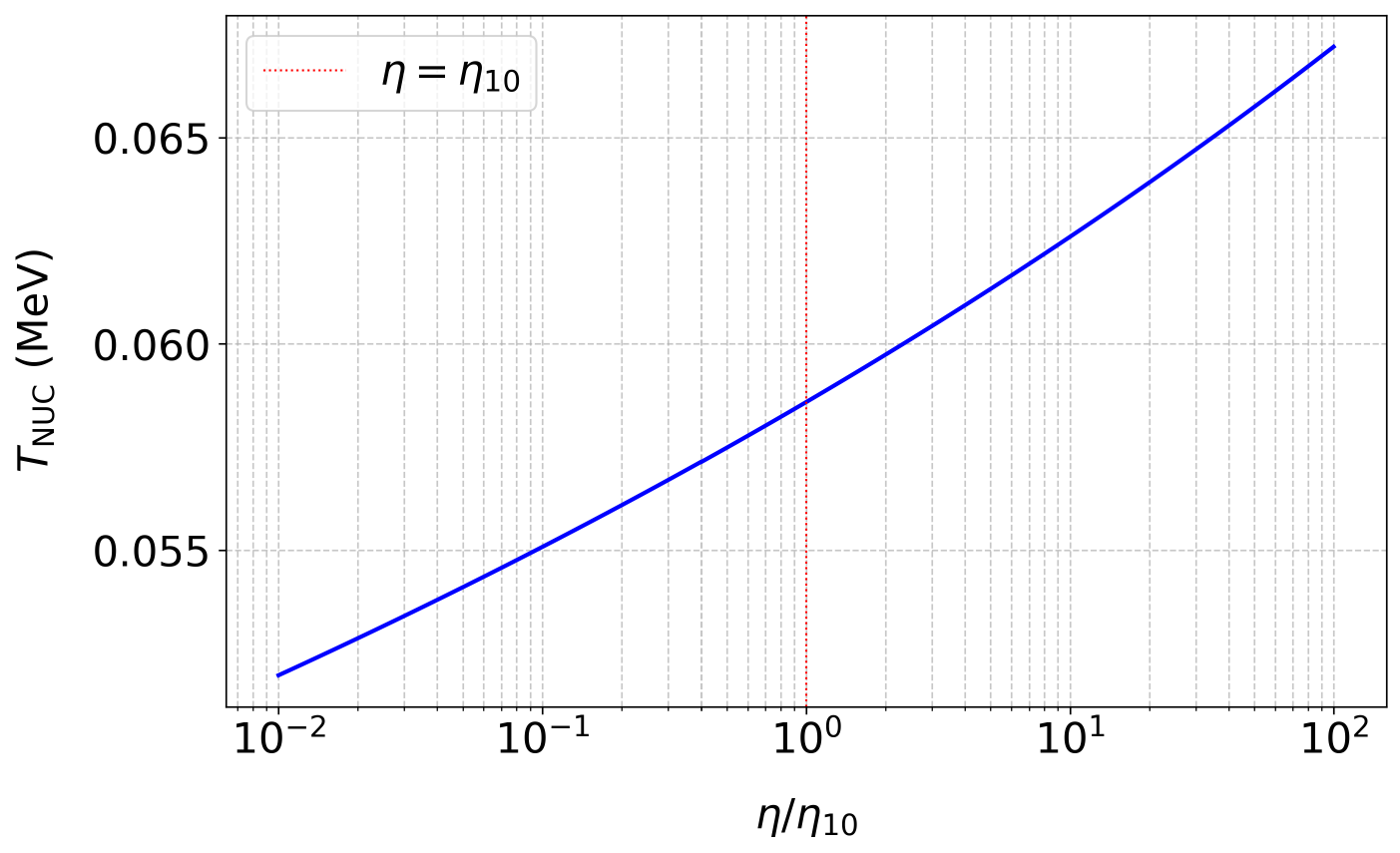}
\caption{The resulting $T_{\text{NUC}}$ as a function of $\eta/\eta_{10}$ for deuterium in the GIMOG model.}
\label{fig:T_NUC}
\end{figure}

Adopting the standard normalization $\eta_{10} = 10^{-10}$, we numerically evaluate the deuterium formation temperature using Eq.~\eqref{eq:T_NUC_approx}. Fig.~\ref{fig:T_NUC} presents the resulting $T_{\text{NUC}}$ as a function of $\eta/\eta_{10}$ across the parameter space $\eta \in (10^{-12} - 10^{-8})$. Our analysis demonstrates that the characteristic temperature scale for the onset of deuterium synthesis lies within $T_{\text{NUC}} \approx (0.05 - 0.07)\,\text{MeV}$ for the considered $\eta$ range. This temperature window corresponds to the nuclear binding energy regime in which the deuterium bottleneck is overcome during Big Bang nucleosynthesis.

\begin{figure}[htbp]
\centering
\includegraphics[width=1\columnwidth]{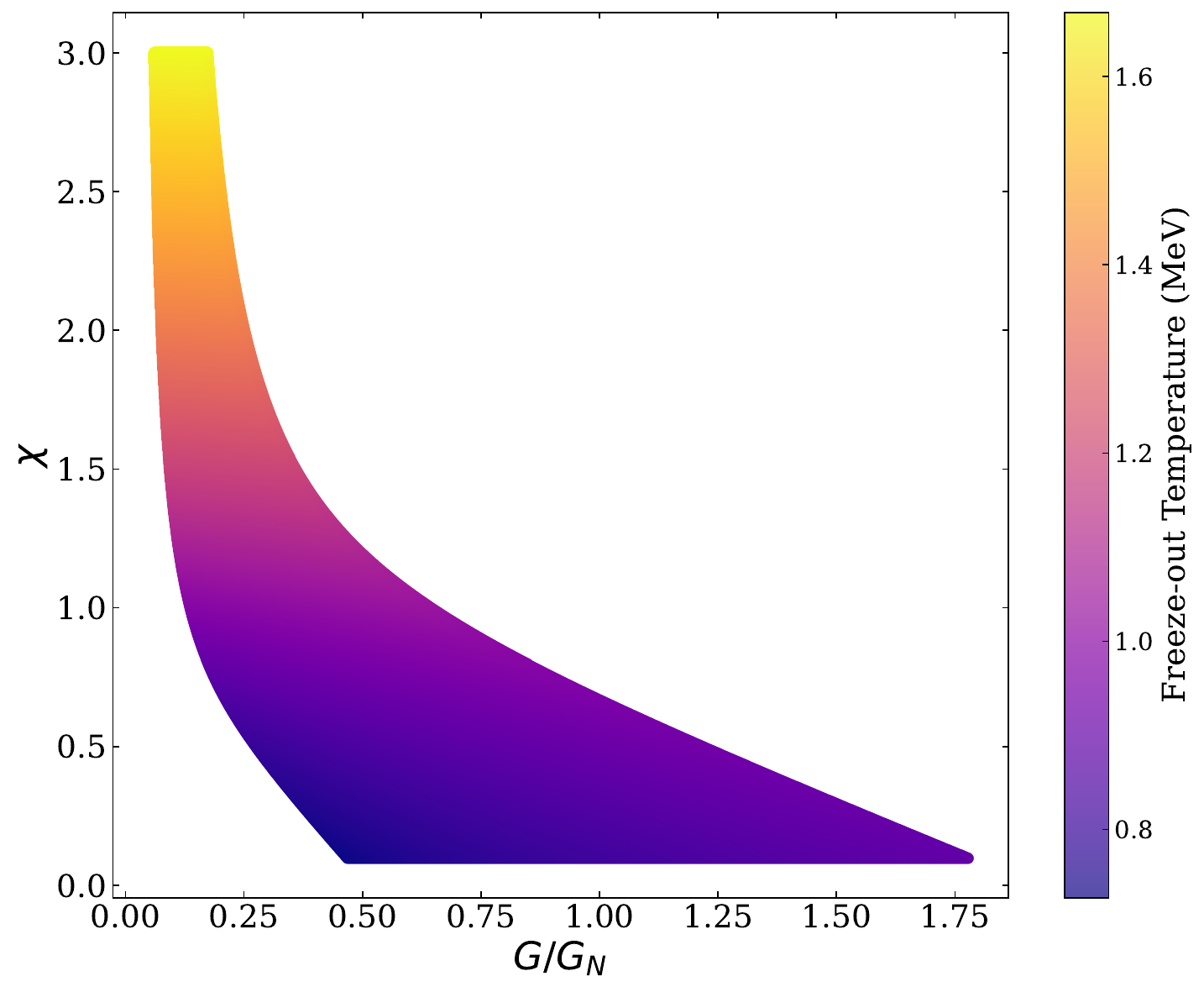}
\caption{Dependence of the dimensionless ratio $\chi$ on $G/G_N$ across varying freeze-out temperatures $T_f$ (in MeV).}
\label{fig:T_f}
\end{figure}

Fig.~\ref{fig:T_f} illustrates $\chi$ as a function of $G/G_N$ for various freeze-out temperatures $T_f$ (expressed in MeV). In the standard cosmological model, characterized by $G/G_N = 1$ and $\chi = 0$, the predictions of general relativity (GR) are successfully recovered. Deviations from GR emerge as either $G$ or $\chi$ increases: an enhanced gravitational constant $G$ induces a moderate increase in $T_f$, whereas a larger value of $\chi$ drives a more substantial shift, elevating $T_f$ significantly. This asymmetry suggests that primordial nucleosynthesis is markedly more sensitive to $\chi$-driven modifications than to variations in $G$ alone.

\begin{figure}[htbp]
\centering
\includegraphics[width=0.95\columnwidth]{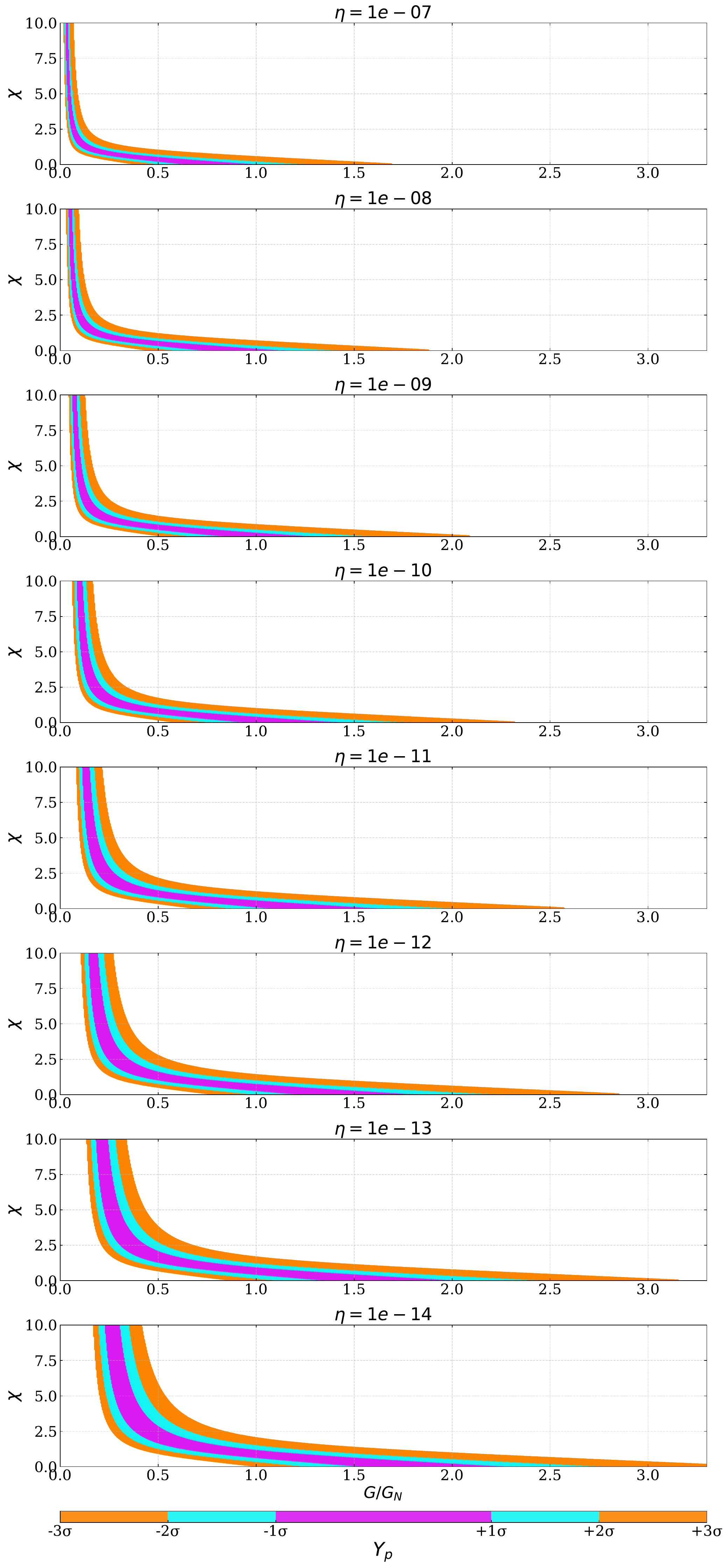}
\caption{Constraints on modified gravity parameters ($\chi$, $G/G_N$) from Big Bang Nucleosynthesis. Colored bands show the $3\sigma$ allowed regions for $Y_p$ at different values of $\eta$.}
\label{fig:eta}
\end{figure}

Furthermore, Fig.~\ref{fig:eta} details the constraints on the dimensionless ratios $\chi$ versus $G/G_N$ derived from Big Bang nucleosynthesis, demarcating the $3\sigma$ confidence regions for the primordial helium-4 abundance ($Y_p$). Each contour corresponds to a distinct baryon-to-photon ratio, $\eta$. As $\eta$ decreases, the permissible parameter space expands significantly, reflecting the heightened sensitivity of light-element production to gravitational modifications in low-baryon-density regimes. The shaded bands denote the theoretical uncertainties inherent in the $Y_p$ calculations.

\begin{figure}[htbp]
\centering
\includegraphics[width=1\columnwidth]{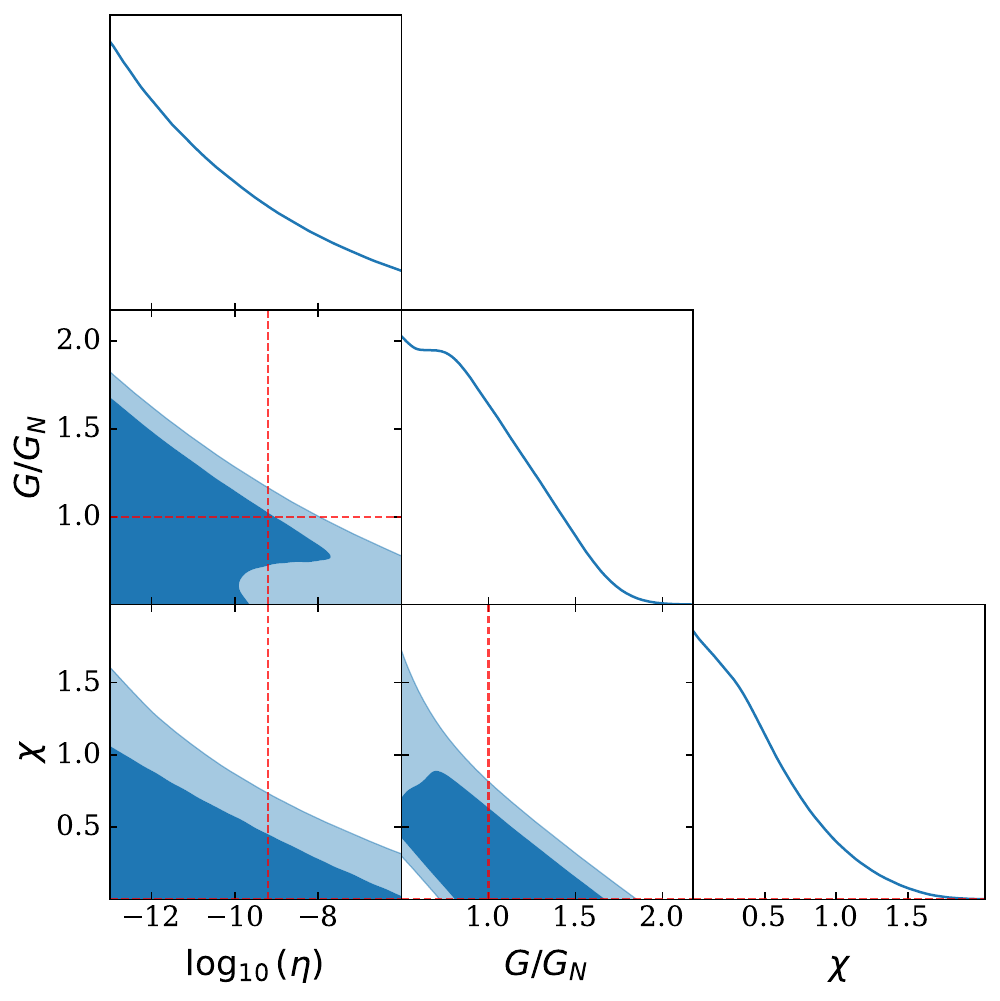}
\caption{68\% and 95\% Bayesian credible regions for $(\log_{10}\eta, G/G_N, \chi)$ from BBN $Y_p$ constraints. Red dashed lines indicate Planck 2018, GR, and $\chi=0$.}
\label{fig:MCMC_BBN}
\end{figure}

Fig.~\ref{fig:MCMC_BBN} presents the marginalized posterior distributions for the key parameters of our modified gravity framework, obtained via a Markov Chain Monte Carlo (MCMC) analysis utilizing the \texttt{GetDist} package. The numerical results are summarized in Table~\ref{tab:bbn_results}.
\begin{table}[htbp]
\centering
\caption{Best-fit parameters and their corresponding $68\%$ confidence limits for the BBN model obtained from the MCMC analysis.}
\label{tab:bbn_results}
\begin{tabular*}{\columnwidth}{@{\extracolsep{\fill}}lcc}
	\hline
	\hline
	Parameter & Best-fit Value & Uncertainty ($1\sigma$) \\
	\hline
	$\log_{10}(\eta)$ & $-10.9085$ & $^{+2.5303}_{-1.5325}$ \\
	$G/G_N$           & $0.8998$   & $^{+0.3878}_{-0.2760}$ \\
	$\chi$            & $0.3860$   & $^{+0.4491}_{-0.2737}$ \\
	\hline
	\hline
\end{tabular*}
\end{table}
\section{Multi-Epoch Constraints on GIMOG Parameters}
\label{sec:multi_epoch}
A stringent validation of any modified gravity theory is its consistency across disparate cosmological epochs. To this end, we conduct a rigorous assessment of the GIMOG model's cosmological viability by comparing constraints on the effective gravitational constant ratio $G/G_N$ derived from two distinct eras: late-time Universe observations (Pantheon+ Type Ia Supernovae) and primordial nucleosynthesis (BBN based on the primordial helium-4 abundance). This multi-epoch analysis critically evaluates the model's capacity to simultaneously account for late-time cosmic acceleration and early-Universe BBN physics within a unified gravitational framework. The complementary nature of late-time and primordial probes provides a robust consistency test that cannot be achieved with either dataset in isolation, highlighting the vital necessity of multi-epoch validation.

As detailed in Sec.~\ref{sec:Sn_method}, the Pantheon+ SNe Ia sample constrains the parameters governing the late-time cosmic expansion ($z \lesssim 2.3$), in particular the effective matter-like coefficient (A). Under the baryon-only assumption and an external prior on $\Omega_{0b}$, this constraint can be translated into an inferred late-time value of $G/G_N$. Conversely, the primordial helium-4 mass fraction ($Y_p$) serves as a precision test of gravitational physics during BBN, providing an independent early-time constraint on the allowed gravitational coupling.

\begin{figure}[htbp]
\centering
\includegraphics[width=1\columnwidth]{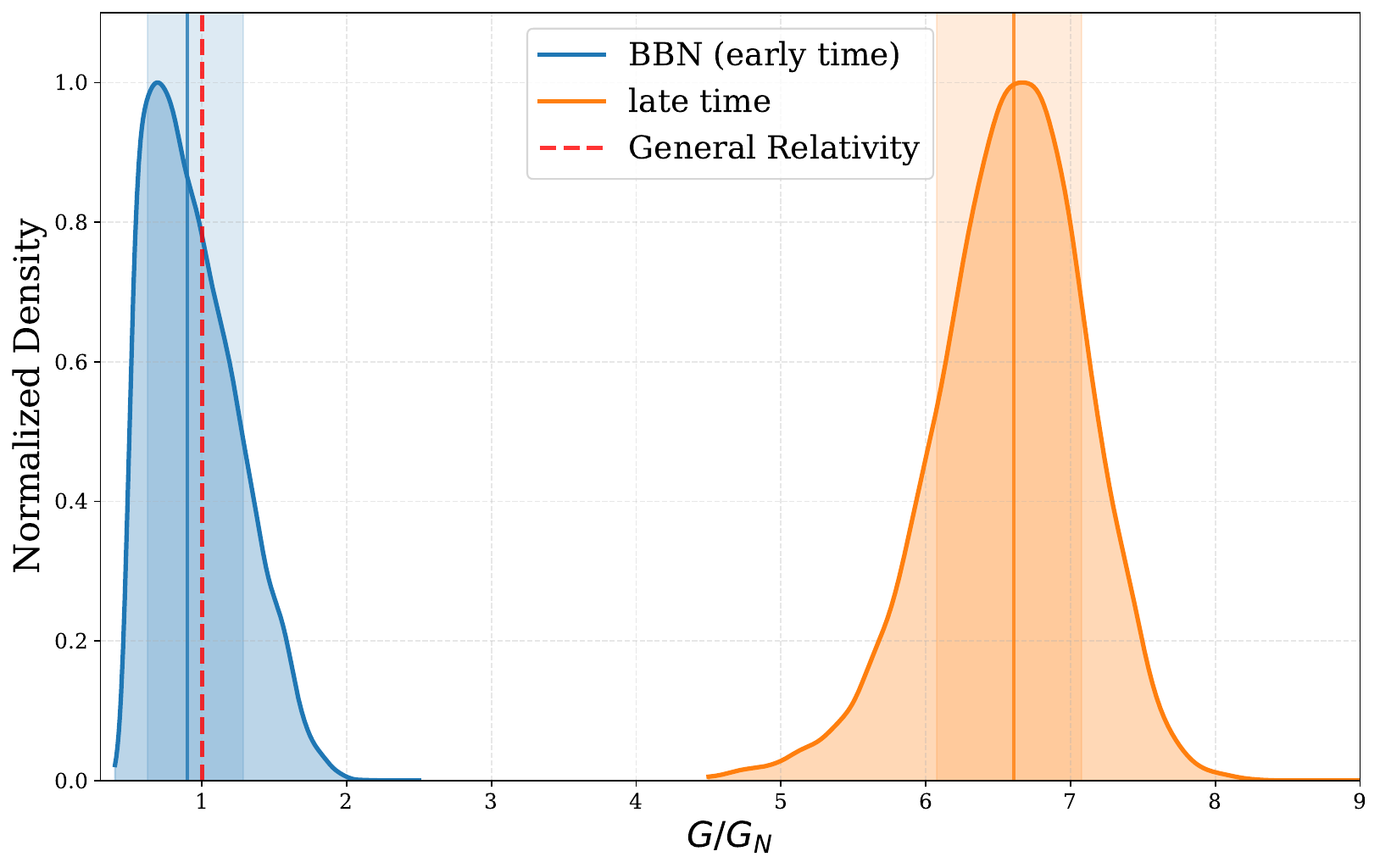}
\caption{Normalized posterior distributions of $G/G_N$ from BBN (early Universe, blue) and Pantheon+ SNe Ia (late Universe, orange). Solid vertical lines indicate the median values, shaded regions show $68\%$ confidence intervals, and the dashed line marks the General Relativity prediction ($G/G_N=1$).}
\label{fig:both_G}
\end{figure}

Fig.~\ref{fig:both_G} displays the normalized posterior distributions for the fundamental parameter $G/G_N$, where the orange curve corresponds to the constraints from SNe Ia (late time) and the blue curve represents the bounds derived from BBN (early time). 

The joint analysis reveals a pronounced tension in the $G/G_N$ constraints derived from the two epochs. The early-time (BBN) measurement is fully consistent with the General Relativity prediction ($G/G_N = 1$) within $1\sigma$ uncertainty, while the late-time (Pantheon+) measurement exhibits a significant deviation, favoring a value approximately six times larger than the Newtonian constant. The $68\%$ confidence intervals of the two posterior distributions do not overlap, pointing toward an epoch-dependent gravitational behavior where early and late observations favor distinct regions of the parameter space.

This incompatibility cannot be resolved within the static formulation of the model evaluated here. Reconciling the primordial and late-time regimes would necessitate an explicitly time-dependent effective gravitational coupling, positing a smooth evolution from $G \approx G_N$ during nucleosynthesis to $G \sim 6\,G_N$ in the late Universe. Although the potential for the $G$ field has been switched off in the present analysis, a fully dynamical mechanism could naturally drive such an evolution.
\section{Conclusion}
\label{sec:conc}

In this paper, we have explored the cosmological implications of a gauge-invariant model of modified gravity (GIMOG), in which gravity is dynamically generated via spontaneous symmetry breaking. We have shown that this framework is capable of reproducing the broad strokes of the Universe's thermal history. It yields a pre-inflationary epoch, transitions into the standard radiation and matter-dominated eras, and can reproduce the observed late-time cosmic acceleration without introducing a fundamental cosmological constant; in this framework, the acceleration is effectively sourced by the scalar-field potential.

To rigorously test the viability of the GIMOG scenario, we confronted the model with both early- and late-Universe observational data. Utilizing the Pantheon+ supernova compilation and cosmic chronometer measurements, we found that the late-time expansion dynamics are well-accommodated, though the data prefer an effective gravitational coupling stronger than the standard Newtonian value. Conversely, we applied Big Bang Nucleosynthesis (BBN) constraints---relying primarily on the primordial $^{4}\mathrm{He}$ abundance---to restrict the parameter space in the early Universe.

Our joint analysis exposes a significant tension when assuming constant model parameters: the parameter space favored by late-time cosmological data is not entirely consistent with the stringent bounds imposed by BBN. To seamlessly connect the early- and late-time dynamics and resolve this discrepancy, our findings indicate the necessity of incorporating a smooth, time-dependent variation of the effective gravitational constant, $G$. 

Looking ahead, exploring the time-evolution of $G$ within this framework remains a critical objective. Future work will extend the GIMOG model to the level of linear cosmological perturbations. By investigating cosmic microwave background (CMB) anisotropies, the growth rate of large-scale structures, and the propagation of gravitational waves, we aim to place more robust, scale-dependent constraints on the theory and further assess its viability as a comprehensive alternative to the standard $\Lambda\text{CDM}$ paradigm.

\section*{Acknowledgements}
This research was supported by Sharif University of Technology's Office of the Vice President for Research under Grant No.~G4010204.

\section*{Data Availability Statement}
The observational datasets analyzed during the current study (including the Pantheon+ Type Ia Supernovae compilation and the Cosmic Chronometers data) are publicly available in established repositories. No new data were generated.

\bibliography{apssamp} 

\begin{thebibliography}{29}%
\makeatletter
\providecommand \@ifxundefined [1]{%
 \@ifx{#1\undefined}
}%
\providecommand \@ifnum [1]{%
 \ifnum #1\expandafter \@firstoftwo
 \else \expandafter \@secondoftwo
 \fi
}%
\providecommand \@ifx [1]{%
 \ifx #1\expandafter \@firstoftwo
 \else \expandafter \@secondoftwo
 \fi
}%
\providecommand \natexlab [1]{#1}%
\providecommand \enquote  [1]{``#1''}%
\providecommand \bibnamefont  [1]{#1}%
\providecommand \bibfnamefont [1]{#1}%
\providecommand \citenamefont [1]{#1}%
\providecommand \href@noop [0]{\@secondoftwo}%
\providecommand \href [0]{\begingroup \@sanitize@url \@href}%
\providecommand \@href[1]{\@@startlink{#1}\@@href}%
\providecommand \@@href[1]{\endgroup#1\@@endlink}%
\providecommand \@sanitize@url [0]{\catcode `\\12\catcode `\$12\catcode
  `\&12\catcode `\#12\catcode `\^12\catcode `\_12\catcode `\%12\relax}%
\providecommand \@@startlink[1]{}%
\providecommand \@@endlink[0]{}%
\providecommand \url  [0]{\begingroup\@sanitize@url \@url }%
\providecommand \@url [1]{\endgroup\@href {#1}{\urlprefix }}%
\providecommand \urlprefix  [0]{URL }%
\providecommand \Eprint [0]{\href }%
\providecommand \doibase [0]{https://doi.org/}%
\providecommand \selectlanguage [0]{\@gobble}%
\providecommand \bibinfo  [0]{\@secondoftwo}%
\providecommand \bibfield  [0]{\@secondoftwo}%
\providecommand \translation [1]{[#1]}%
\providecommand \BibitemOpen [0]{}%
\providecommand \bibitemStop [0]{}%
\providecommand \bibitemNoStop [0]{.\EOS\space}%
\providecommand \EOS [0]{\spacefactor3000\relax}%
\providecommand \BibitemShut  [1]{\csname bibitem#1\endcsname}%
\let\auto@bib@innerbib\@empty
\bibitem [{\citenamefont {Rubin}\ and\ \citenamefont
  {Ford~Jr}(1970)}]{rubin1970rotation}%
  \BibitemOpen
  \bibfield  {author} {\bibinfo {author} {\bibfnamefont {V.~C.}\ \bibnamefont
  {Rubin}}\ and\ \bibinfo {author} {\bibfnamefont {W.~K.}\ \bibnamefont
  {Ford~Jr}},\ }\bibfield  {title} {\bibinfo {title} {Rotation of the andromeda
  nebula from a spectroscopic survey of emission regions},\ }\href@noop {}
  {\bibfield  {journal} {\bibinfo  {journal} {Astrophysical Journal, vol. 159,
  p. 379}\ }\textbf {\bibinfo {volume} {159}},\ \bibinfo {pages} {379}
  (\bibinfo {year} {1970})}\BibitemShut {NoStop}%
\bibitem [{\citenamefont {Rubin}(2000)}]{rubin2000one}%
  \BibitemOpen
  \bibfield  {author} {\bibinfo {author} {\bibfnamefont {V.~C.}\ \bibnamefont
  {Rubin}},\ }\bibfield  {title} {\bibinfo {title} {One hundred years of
  rotating galaxies},\ }\href@noop {} {\bibfield  {journal} {\bibinfo
  {journal} {Publications of the Astronomical Society of the Pacific}\ }\textbf
  {\bibinfo {volume} {112}},\ \bibinfo {pages} {747} (\bibinfo {year}
  {2000})}\BibitemShut {NoStop}%
\bibitem [{\citenamefont {Sofue}\ and\ \citenamefont
  {Rubin}(2001)}]{sofue2001rotation}%
  \BibitemOpen
  \bibfield  {author} {\bibinfo {author} {\bibfnamefont {Y.}~\bibnamefont
  {Sofue}}\ and\ \bibinfo {author} {\bibfnamefont {V.}~\bibnamefont {Rubin}},\
  }\bibfield  {title} {\bibinfo {title} {Rotation curves of spiral galaxies},\
  }\href@noop {} {\bibfield  {journal} {\bibinfo  {journal} {Annual Review of
  Astronomy and Astrophysics}\ }\textbf {\bibinfo {volume} {39}},\ \bibinfo
  {pages} {137} (\bibinfo {year} {2001})}\BibitemShut {NoStop}%
\bibitem [{\citenamefont {Carlberg}\ \emph {et~al.}(1996)\citenamefont
  {Carlberg} \emph {et~al.}}]{carlberg1996galaxy}%
  \BibitemOpen
  \bibfield  {author} {\bibinfo {author} {\bibfnamefont {R.}~\bibnamefont
  {Carlberg}} \emph {et~al.},\ }\bibfield  {title} {\bibinfo {title} {Galaxy
  cluster virial masses and omega},\ }\href@noop {} {\bibfield  {journal}
  {\bibinfo  {journal} {Astrophysical Journal v. 462, p. 32}\ }\textbf
  {\bibinfo {volume} {462}},\ \bibinfo {pages} {32} (\bibinfo {year}
  {1996})}\BibitemShut {NoStop}%
\bibitem [{\citenamefont {Bartelmann}\ and\ \citenamefont
  {Schneider}(2001)}]{bartelmann2001weak}%
  \BibitemOpen
  \bibfield  {author} {\bibinfo {author} {\bibfnamefont {M.}~\bibnamefont
  {Bartelmann}}\ and\ \bibinfo {author} {\bibfnamefont {P.}~\bibnamefont
  {Schneider}},\ }\bibfield  {title} {\bibinfo {title} {Weak gravitational
  lensing},\ }\href@noop {} {\bibfield  {journal} {\bibinfo  {journal} {Physics
  Reports}\ }\textbf {\bibinfo {volume} {340}},\ \bibinfo {pages} {291}
  (\bibinfo {year} {2001})}\BibitemShut {NoStop}%
\bibitem [{\citenamefont {Arcadi}\ \emph {et~al.}(2025)\citenamefont {Arcadi},
  \emph {et~al.}}]{arcadi2025waning}%
  \BibitemOpen
  \bibfield  {author} {\bibinfo {author} {\bibfnamefont {G.}~\bibnamefont
  {Arcadi}}, , \emph {et~al.},\ }\bibfield  {title} {\bibinfo {title} {The
  waning of the wimp: endgame?},\ }\href@noop {} {\bibfield  {journal}
  {\bibinfo  {journal} {The European Physical Journal C}\ }\textbf {\bibinfo
  {volume} {85}},\ \bibinfo {pages} {152} (\bibinfo {year} {2025})}\BibitemShut
  {NoStop}%
\bibitem [{\citenamefont {Kuster}\ \emph {et~al.}(2007)\citenamefont {Kuster}
  \emph {et~al.}}]{kuster2007axions}%
  \BibitemOpen
  \bibfield  {author} {\bibinfo {author} {\bibfnamefont {M.}~\bibnamefont
  {Kuster}} \emph {et~al.},\ }\href@noop {} {\emph {\bibinfo {title} {Axions:
  Theory, cosmology, and experimental searches}}},\ Vol.\ \bibinfo {volume}
  {741}\ (\bibinfo  {publisher} {Springer Science \& Business Media},\ \bibinfo
  {year} {2007})\BibitemShut {NoStop}%
\bibitem [{\citenamefont {Milgrom}(1988)}]{milgrom1988use}%
  \BibitemOpen
  \bibfield  {author} {\bibinfo {author} {\bibfnamefont {M.}~\bibnamefont
  {Milgrom}},\ }\bibfield  {title} {\bibinfo {title} {On the use of galaxy
  rotation curves to test the modified dynamics},\ }\href@noop {} {\bibfield
  {journal} {\bibinfo  {journal} {Astrophysical Journal}\ }\textbf {\bibinfo
  {volume} {333}},\ \bibinfo {pages} {689} (\bibinfo {year}
  {1988})}\BibitemShut {NoStop}%
\bibitem [{\citenamefont {Moffat}(2006)}]{moffat2006scalar}%
  \BibitemOpen
  \bibfield  {author} {\bibinfo {author} {\bibfnamefont {J.~W.}\ \bibnamefont
  {Moffat}},\ }\bibfield  {title} {\bibinfo {title} {Scalar--tensor--vector
  gravity theory},\ }\href@noop {} {\bibfield  {journal} {\bibinfo  {journal}
  {Journal of Cosmology and Astroparticle Physics}\ }\textbf {\bibinfo {volume}
  {2006}}\bibinfo  {number} { (03)},\ \bibinfo {pages} {004}}\BibitemShut
  {NoStop}%
\bibitem [{\citenamefont {Moffat}\ and\ \citenamefont
  {Rahvar}(2013)}]{moffat2013mog}%
  \BibitemOpen
\bibfield  {number} {  }\bibfield  {author} {\bibinfo {author} {\bibfnamefont
  {J.}~\bibnamefont {Moffat}}\ and\ \bibinfo {author} {\bibfnamefont
  {S.}~\bibnamefont {Rahvar}},\ }\bibfield  {title} {\bibinfo {title} {The mog
  weak field approximation and observational test of galaxy rotation curves},\
  }\href@noop {} {\bibfield  {journal} {\bibinfo  {journal} {Monthly Notices of
  the Royal Astronomical Society}\ }\textbf {\bibinfo {volume} {436}},\
  \bibinfo {pages} {1439} (\bibinfo {year} {2013})}\BibitemShut {NoStop}%
\bibitem [{\citenamefont {Moffat}\ and\ \citenamefont
  {Rahvar}(2014)}]{moffat2014mog}%
  \BibitemOpen
  \bibfield  {author} {\bibinfo {author} {\bibfnamefont {J.}~\bibnamefont
  {Moffat}}\ and\ \bibinfo {author} {\bibfnamefont {S.}~\bibnamefont
  {Rahvar}},\ }\bibfield  {title} {\bibinfo {title} {The mog weak field
  approximation--ii. observational test of chandra x-ray clusters},\
  }\href@noop {} {\bibfield  {journal} {\bibinfo  {journal} {Monthly Notices of
  the Royal Astronomical Society}\ }\textbf {\bibinfo {volume} {441}},\
  \bibinfo {pages} {3724} (\bibinfo {year} {2014})}\BibitemShut {NoStop}%
\bibitem [{\citenamefont {Rahvar}\ and\ \citenamefont
  {Moffat}(2019)}]{rahvar2019propagation}%
  \BibitemOpen
  \bibfield  {author} {\bibinfo {author} {\bibfnamefont {S.}~\bibnamefont
  {Rahvar}}\ and\ \bibinfo {author} {\bibfnamefont {J.}~\bibnamefont
  {Moffat}},\ }\bibfield  {title} {\bibinfo {title} {Propagation of
  electromagnetic waves in mog: gravitational lensing},\ }\href@noop {}
  {\bibfield  {journal} {\bibinfo  {journal} {Monthly Notices of the Royal
  Astronomical Society}\ }\textbf {\bibinfo {volume} {482}},\ \bibinfo {pages}
  {4514} (\bibinfo {year} {2019})}\BibitemShut {NoStop}%
\bibitem [{\citenamefont {Rahvar}(2022)}]{rahvar2022hamiltonian}%
  \BibitemOpen
  \bibfield  {author} {\bibinfo {author} {\bibfnamefont {S.}~\bibnamefont
  {Rahvar}},\ }\bibfield  {title} {\bibinfo {title} {Hamiltonian formalism for
  dynamics of particles in mog},\ }\href@noop {} {\bibfield  {journal}
  {\bibinfo  {journal} {Monthly Notices of the Royal Astronomical Society}\
  }\textbf {\bibinfo {volume} {514}},\ \bibinfo {pages} {4601} (\bibinfo {year}
  {2022})}\BibitemShut {NoStop}%
\bibitem [{\citenamefont {Rouhani}\ and\ \citenamefont
  {Rahvar}(2024)}]{rouhani2024mog}%
  \BibitemOpen
  \bibfield  {author} {\bibinfo {author} {\bibfnamefont {S.}~\bibnamefont
  {Rouhani}}\ and\ \bibinfo {author} {\bibfnamefont {S.}~\bibnamefont
  {Rahvar}},\ }\bibfield  {title} {\bibinfo {title} {Mog as symmetry breaking
  in scalar--vector--tensor gravity},\ }\href@noop {} {\bibfield  {journal}
  {\bibinfo  {journal} {Monthly Notices of the Royal Astronomical Society}\
  }\textbf {\bibinfo {volume} {527}},\ \bibinfo {pages} {2831} (\bibinfo {year}
  {2024})}\BibitemShut {NoStop}%
\bibitem [{\citenamefont {Scolnic}\ \emph
  {et~al.}(2022{\natexlab{a}})\citenamefont {Scolnic} \emph
  {et~al.}}]{PantheonData}%
  \BibitemOpen
  \bibfield  {author} {\bibinfo {author} {\bibfnamefont {D.~M.}\ \bibnamefont
  {Scolnic}} \emph {et~al.},\ }\href@noop {} {\bibinfo {title} {{Pantheon}+
  {SH0ES} {D}ata {R}elease}},\ \bibinfo {howpublished}
  {\url{https://github.com/PantheonPlusSH0ES/DataRelease}} (\bibinfo {year}
  {2022}{\natexlab{a}}),\ \bibinfo {note} {accessed: 2025-05-10}\BibitemShut
  {NoStop}%
\bibitem [{\citenamefont {Scolnic}\ \emph
  {et~al.}(2022{\natexlab{b}})\citenamefont {Scolnic} \emph
  {et~al.}}]{scolnic2022pantheon+}%
  \BibitemOpen
  \bibfield  {author} {\bibinfo {author} {\bibfnamefont {D.}~\bibnamefont
  {Scolnic}} \emph {et~al.},\ }\bibfield  {title} {\bibinfo {title} {The
  pantheon+ analysis: the full data set and light-curve release},\ }\href@noop
  {} {\bibfield  {journal} {\bibinfo  {journal} {The Astrophysical Journal}\
  }\textbf {\bibinfo {volume} {938}},\ \bibinfo {pages} {113} (\bibinfo {year}
  {2022}{\natexlab{b}})}\BibitemShut {NoStop}%
\bibitem [{\citenamefont {Jimenez}\ and\ \citenamefont
  {Loeb}(2002)}]{jimenez2002constraining}%
  \BibitemOpen
  \bibfield  {author} {\bibinfo {author} {\bibfnamefont {R.}~\bibnamefont
  {Jimenez}}\ and\ \bibinfo {author} {\bibfnamefont {A.}~\bibnamefont {Loeb}},\
  }\bibfield  {title} {\bibinfo {title} {Constraining cosmological parameters
  based on relative galaxy ages},\ }\href@noop {} {\bibfield  {journal}
  {\bibinfo  {journal} {The Astrophysical Journal}\ }\textbf {\bibinfo {volume}
  {573}},\ \bibinfo {pages} {37} (\bibinfo {year} {2002})}\BibitemShut
  {NoStop}%
\bibitem [{\citenamefont {Simon}\ \emph {et~al.}(2005)\citenamefont {Simon},
  \citenamefont {Verde},\ and\ \citenamefont {Jimenez}}]{simon2005constraints}%
  \BibitemOpen
  \bibfield  {author} {\bibinfo {author} {\bibfnamefont {J.}~\bibnamefont
  {Simon}}, \bibinfo {author} {\bibfnamefont {L.}~\bibnamefont {Verde}},\ and\
  \bibinfo {author} {\bibfnamefont {R.}~\bibnamefont {Jimenez}},\ }\bibfield
  {title} {\bibinfo {title} {Constraints on the redshift dependence of the dark
  energy potential},\ }\href@noop {} {\bibfield  {journal} {\bibinfo  {journal}
  {Physical Review D—Particles, Fields, Gravitation, and Cosmology}\ }\textbf
  {\bibinfo {volume} {71}},\ \bibinfo {pages} {123001} (\bibinfo {year}
  {2005})}\BibitemShut {NoStop}%
\bibitem [{\citenamefont {Moresco}\ \emph {et~al.}(2016)\citenamefont {Moresco}
  \emph {et~al.}}]{moresco20166}%
  \BibitemOpen
  \bibfield  {author} {\bibinfo {author} {\bibfnamefont {M.}~\bibnamefont
  {Moresco}} \emph {et~al.},\ }\bibfield  {title} {\bibinfo {title} {A 6\%
  measurement of the {H}ubble parameter at $z \sim 0.45$: direct evidence of
  the epoch of cosmic re-acceleration},\ }\href@noop {} {\bibfield  {journal}
  {\bibinfo  {journal} {Journal of Cosmology and Astroparticle Physics}\
  }\textbf {\bibinfo {volume} {2016}}\bibinfo  {number} { (05)},\ \bibinfo
  {pages} {014}}\BibitemShut {NoStop}%
\bibitem [{\citenamefont {Negrelli}\ \emph {et~al.}(2020)\citenamefont
  {Negrelli}, \citenamefont {Kraiselburd}, \citenamefont {Landau},\ and\
  \citenamefont {Scoccola}}]{negrelli2020testing}%
  \BibitemOpen
\bibfield  {number} {  }\bibfield  {author} {\bibinfo {author} {\bibfnamefont
  {C.}~\bibnamefont {Negrelli}}, \bibinfo {author} {\bibfnamefont
  {L.}~\bibnamefont {Kraiselburd}}, \bibinfo {author} {\bibfnamefont
  {S.}~\bibnamefont {Landau}},\ and\ \bibinfo {author} {\bibfnamefont {C.~G.}\
  \bibnamefont {Scoccola}},\ }\bibfield  {title} {\bibinfo {title} {Testing
  modified gravity theory (mog) with type ia supernovae, cosmic chronometers
  and baryon acoustic oscillations},\ }\href@noop {} {\bibfield  {journal}
  {\bibinfo  {journal} {Journal of Cosmology and Astroparticle Physics}\
  }\textbf {\bibinfo {volume} {2020}}\bibinfo  {number} { (07)},\ \bibinfo
  {pages} {015}}\BibitemShut {NoStop}%
\bibitem [{\citenamefont {Riess}\ \emph {et~al.}(2022)\citenamefont {Riess}
  \emph {et~al.}}]{riess2022comprehensive}%
  \BibitemOpen
\bibfield  {number} {  }\bibfield  {author} {\bibinfo {author} {\bibfnamefont
  {A.~G.}\ \bibnamefont {Riess}} \emph {et~al.},\ }\bibfield  {title} {\bibinfo
  {title} {A {C}omprehensive {M}easurement of the {L}ocal {V}alue of the
  {H}ubble {C}onstant with 1~km~s$^{-1}$~{Mpc}$^{-1}$ {U}ncertainty from the
  {Hubble} {Space} {T}elescope and the {SH0ES} {T}eam},\ }\href@noop {}
  {\bibfield  {journal} {\bibinfo  {journal} {The Astrophysical journal
  letters}\ }\textbf {\bibinfo {volume} {934}},\ \bibinfo {pages} {L7}
  (\bibinfo {year} {2022})}\BibitemShut {NoStop}%
\bibitem [{\citenamefont {Adame}\ \emph {et~al.}(2025)\citenamefont {Adame}
  \emph {et~al.}}]{adame2025desi}%
  \BibitemOpen
  \bibfield  {author} {\bibinfo {author} {\bibfnamefont {A.}~\bibnamefont
  {Adame}} \emph {et~al.},\ }\bibfield  {title} {\bibinfo {title} {Desi 2024 v:
  Full-shape galaxy clustering from galaxies and quasars},\ }\href@noop {}
  {\bibfield  {journal} {\bibinfo  {journal} {Journal of Cosmology and
  Astroparticle Physics}\ }\textbf {\bibinfo {volume} {2025}}\bibinfo  {number}
  { (09)},\ \bibinfo {pages} {008}}\BibitemShut {NoStop}%
\bibitem [{\citenamefont {Padmanabhan}(1993)}]{padmanabhan1993structure}%
  \BibitemOpen
\bibfield  {number} {  }\bibfield  {author} {\bibinfo {author} {\bibfnamefont
  {T.}~\bibnamefont {Padmanabhan}},\ }\href@noop {} {\emph {\bibinfo {title}
  {Structure formation in the universe}}}\ (\bibinfo  {publisher} {Cambridge
  University Press},\ \bibinfo {year} {1993})\BibitemShut {NoStop}%
\bibitem [{\citenamefont {Kolb}(2018)}]{kolb2018early}%
  \BibitemOpen
  \bibfield  {author} {\bibinfo {author} {\bibfnamefont {E.}~\bibnamefont
  {Kolb}},\ }\href@noop {} {\emph {\bibinfo {title} {The early universe}}}\
  (\bibinfo  {publisher} {CRC press},\ \bibinfo {year} {2018})\BibitemShut
  {NoStop}%
\bibitem [{\citenamefont {Moffat}\ and\ \citenamefont
  {Toth}(2009)}]{moffat2009fundamental}%
  \BibitemOpen
  \bibfield  {author} {\bibinfo {author} {\bibfnamefont {J.}~\bibnamefont
  {Moffat}}\ and\ \bibinfo {author} {\bibfnamefont {V.~T.}\ \bibnamefont
  {Toth}},\ }\bibfield  {title} {\bibinfo {title} {Fundamental parameter-free
  solutions in modified gravity},\ }\href@noop {} {\bibfield  {journal}
  {\bibinfo  {journal} {Classical and Quantum Gravity}\ }\textbf {\bibinfo
  {volume} {26}},\ \bibinfo {pages} {085002} (\bibinfo {year}
  {2009})}\BibitemShut {NoStop}%
\bibitem [{\citenamefont {Baumann}(2022)}]{baumann2022cosmology}%
  \BibitemOpen
  \bibfield  {author} {\bibinfo {author} {\bibfnamefont {D.}~\bibnamefont
  {Baumann}},\ }\href@noop {} {\emph {\bibinfo {title} {Cosmology}}}\ (\bibinfo
   {publisher} {Cambridge University Press},\ \bibinfo {year}
  {2022})\BibitemShut {NoStop}%
\bibitem [{\citenamefont {Dodelson}\ and\ \citenamefont
  {Schmidt}(2024)}]{dodelson2024modern}%
  \BibitemOpen
  \bibfield  {author} {\bibinfo {author} {\bibfnamefont {S.}~\bibnamefont
  {Dodelson}}\ and\ \bibinfo {author} {\bibfnamefont {F.}~\bibnamefont
  {Schmidt}},\ }\href@noop {} {\emph {\bibinfo {title} {Modern cosmology}}}\
  (\bibinfo  {publisher} {Elsevier},\ \bibinfo {year} {2024})\BibitemShut
  {NoStop}%
\bibitem [{\citenamefont {Valerdi}\ \emph {et~al.}(2019)\citenamefont
  {Valerdi}, \citenamefont {Peimbert}, \citenamefont {Peimbert},\ and\
  \citenamefont {Sixtos}}]{valerdi2019determination}%
  \BibitemOpen
  \bibfield  {author} {\bibinfo {author} {\bibfnamefont {M.}~\bibnamefont
  {Valerdi}}, \bibinfo {author} {\bibfnamefont {A.}~\bibnamefont {Peimbert}},
  \bibinfo {author} {\bibfnamefont {M.}~\bibnamefont {Peimbert}},\ and\
  \bibinfo {author} {\bibfnamefont {A.}~\bibnamefont {Sixtos}},\ }\bibfield
  {title} {\bibinfo {title} {Determination of the primordial helium abundance
  based on ngc 346, an h ii region of the small magellanic cloud},\ }\href@noop
  {} {\bibfield  {journal} {\bibinfo  {journal} {The Astrophysical Journal}\
  }\textbf {\bibinfo {volume} {876}},\ \bibinfo {pages} {98} (\bibinfo {year}
  {2019})}\BibitemShut {NoStop}%
\bibitem [{\citenamefont {Kolb}\ \emph {et~al.}(1991)\citenamefont {Kolb},
  \citenamefont {Turner},\ and\ \citenamefont {Silk}}]{kolb1991early}%
  \BibitemOpen
  \bibfield  {author} {\bibinfo {author} {\bibfnamefont {E.~W.}\ \bibnamefont
  {Kolb}}, \bibinfo {author} {\bibfnamefont {M.~S.}\ \bibnamefont {Turner}},\
  and\ \bibinfo {author} {\bibfnamefont {J.}~\bibnamefont {Silk}},\ }\href@noop
  {} {\emph {\bibinfo {title} {The {E}arly {U}niverse}}}\ (\bibinfo
  {publisher} {American Institute of Physics},\ \bibinfo {year}
  {1991})\BibitemShut {NoStop}%
\end{thebibliography}%

\end{document}